\documentclass[cmp]{svjour}
\usepackage{graphicx}
\usepackage{mathptmx}
\usepackage{amsmath}
\usepackage{amssymb}
\newtheorem{deft}{Definition}
\newtheorem{thm}{Theorem}
\newtheorem{lem}{Lemma}
\journalname{Communications in Mathematical Physics}

\begin{document}
\title{Density Functional Theory for Hard Particles in N Dimensions}
\author{Stephan Korden}
\institute{Institute of Technical Thermodynamics, RWTH Aachen
University, Schinkelstra\ss e 8, 52062 Aachen, Germany
\email{stephan.korden@rwth-aachen.de}}
\date{Version: \today}
\maketitle

\begin{abstract}
Recently it has been shown that the heuristic Rosenfeld functional derives from the virial expansion for particles 
which overlap in one center. Here, we generalize this approach to any number of intersections. Starting from the 
virial expansion in Ree-Hoover diagrams, it is shown in the first part that each intersection pattern defines exactly 
one infinite class of diagrams. Determining their automorphism groups, we sum over all its elements and derive a 
generic functional. The second part proves that this functional factorizes into a convolute of integral kernels for 
each intersection center. We derive this kernel for N dimensional particles in the N dimensional, flat Euclidean space. 
The third part focuses on three dimensions and determines the functionals for up to four intersection centers, 
comparing the leading order to Rosenfeld's result. We close by proving a generalized form of the Blaschke, Santalo, 
Chern equation of integral geometry.
\end{abstract}

\section{Introduction}\label{intro}
Density functional theory (DFT) is a cornerstone of many-particle physics \cite{gross,evans-dft}. But as of today, no 
systematic derivation for its functionals is known, neither in quantum nor in classical mechanics, turning its 
construction into a skill guided by symmetry considerations and perturbation theory. In the following, we will show 
that classical hard particles provide an exception to this rule.

Generally, a first approximation in the theory of inhomogeneous fluids is the splitting of the interaction potential 
into a soft contribution and an infinitely repulsive part, defining the particle's geometry and dominating the high 
density phase of soft matter \cite{mcdonald}. Because of its singular potential, the hard-particle reference state 
cannot be treated perturbatively. Therefore, it was an important result, when Rosenfeld presented a hard-sphere 
functional based on three observations: the decoupling of the Mayer function into a convolute of single-particle weight 
functions, the scaled-particle differential equation, and the overall scaling property of the free energy 
\cite{rosenfeld-structure,rosenfeld-freezing,rosenfeld2,rosenfeld-mixture}.

Tarazona later showed the inconsistency of this first functional under restriction of the volume to layers, tubes, and 
single-particle caverns \cite{tarazona,tarazona-rosenfeld}. But by adding further terms and two adjusted coefficients, 
the improved functional accurately described the phase diagram of mono-sized hard spheres. Following this strategy, more 
functionals have been proposed \cite{white-bear-1,white-bear-2,santos-1,santos-2} and successfully applied to several 
systems displaying nematic and crystalline phases 
\cite{schmidt-dft,schmidt-mixtures,cros-ros-1,cros-ros-3,mecke-fmt,mecke-fmt2,marechal-1,marechal-2,marechal-3}. 
Nevertheless, predictions of mixtures and systems of more complex geometries still remain limited to low densities.

Despite its success for spheres, the dependence of the fundamental measure theory (FMT) on the scaled particle 
equation, necessary for the non-perturbative map between low and high density, reduces the approach to a heuristic 
framework not extendable to higher order corrections. A first attempt to generalize the approach to further geometries 
than spheres has been taken by Wertheim, relating Rosenfeld's splitting of the Mayer function to the Gauss-Bonnet 
equation 
\cite{wertheim-1,wertheim-2,wertheim-3}. By calculating the third virial coefficient for ellipsoidal particles, he 
demonstrated the excellent agreement with numerical results, thus showing that the 'lost cases' of the Rosenfeld 
functional are a consequence of the single intersection center approximation \cite{wertheim-4}.

Subsequently, the occurrence of the Gauss-Bonnet equation has been explained to be a consequence of the Blaschke, 
Santalo, Chern equation of integral geometry, whose generalization to arbitrary numbers of particles reproduces the 
weight densities entering in Rosenfeld's functional \cite{korden-2}. In combination with the virial expansion in Mayer 
diagrams, it also explained the polynomial structure of the functional and thus eliminated the dependence on the 
empirical scaled particle theory. It also showed that FMT actually defines an expansion of the exact functional 
in the number, or network, of intersection centers. 

The current article completes this investigation. Starting with Sec.~\ref{sec:virial}, it is shown that each 
intersecting network corresponds to a specific class of Ree-Hoover diagrams. Summing over all graphs of a class and 
determining their combinatorial prefactors, yields a generic functional, whose integrant factorizes into a convolute of 
integral kernels, one for each intersection center. The explicit form of this kernel is derived in 
Sec.~\ref{sec:r-functional}, generalizing the Blaschke, Santalo, Chern equation in App.~\ref{sec:appendix_a}. We close 
by presenting examples in Sec.~\ref{sec:rose-n}.
\section{The Virial Expansion}\label{sec:virial}
The decoupling of the Mayer function of hard particles into a convolute of 1-particle weight function dictates the 
further structure of the FMT functional which is independent of geometry and dimensions. This is discussed in 
Sec.~\ref{subsec:intersects} and motivates a change from particle to intersection coordinates. In 
Sec.~\ref{subsec:rh-diagrams} it is shown that this transformation requires a similar change of the virial expansion 
from Mayer to Ree-Hoover diagrams.
\subsection{FMT as an expansion in intersection centers}\label{subsec:intersects}
Let us consider a set of $i=1,\ldots, N$ particles $\Sigma_i$ imbedded into the finite subset $V$ of the flat, 
Euclidean space $D_i: \Sigma_i \hookrightarrow V \subset \mathbb{R}^n$. In the thermodynamic limit $N,V \to \infty$, the 
particle density $\rho = N/V$ is kept constant, while the free energy $F(N,V,T)$ at temperature $T$ becomes the 
function $F(\rho, T)$. For more than one type of particle, the free energy generalizes correspondingly to 
$F(\{\rho_k\}, T)$ for a mixture of $k=1,\ldots, M$ compounds.

Following Hohenberg, Kohn, and Mermin, the thermodynamic equilibrium is defined as the minimum of the positive definite 
grand-canonical free energy $\Omega([\{\mu_k\}], T)$, a functional of the chemical potentials $\mu_k$ and the 
temperature $T$
\begin{equation}\label{hohenberg-kohn}
\Omega([\{\mu_k\}], T) \geq \Omega([\{\mu^{(0)}_k\}], T) \geq 0\;,
\end{equation}
where $\mu^{(0)}_k$ indicates the chemical potential at equilibrium state \cite{mcdonald}. Taking into account possible 
external potentials $\phi_k(\vec r)$ at $\vec r \in \mathbb{R}^n$, the grand-canonical potential is related to the 
free-energy functional by the Legendre transformation in the local 1-particle densities $\rho_k(\vec r)$:
\begin{equation}\label{functional}
\Omega([\{\rho_k\}], T) = F([\{\rho_k\}], T) + \sum_{k=1}^M \int \rho_k(\vec r)(\phi_k(\vec r) - \mu_k)\,d\gamma\;,
\end{equation}
where we introduced the abbreviation
\begin{equation}\label{measure}
\begin{split}
\Gamma(D) &:= \{\gamma = (\vec r, \vec \omega )\;|\; \vec r\in D, \vec \omega \in \text{SO}(n)\}\\
&\hspace{2em} d\gamma_i  = d^nr_i\,d^{\frac{1}{2}n(n-1)}\omega_i
\end{split}
\end{equation}
for the differential volume element of the Euclidean group $\mathbb{E}^n = \mathbb{R}^n\ltimes\text{SO}(n)$. 

However, the fundamental problem of the DFT approach is that, although the Hohenberg-Kohn theorem assures an almost 
unique relationship between interaction potential and free energy, it provides no hint for its derivation. It was 
therefore surprising, when the Rosenfeld functional could be derived from the virial expansion alone \cite{korden-2}.

Substituting the Boltzmann function $e_{ij}$ by Mayer's f-function $f_{ij} = e_{ij} - 1$ in the configuration integral 
and expanding the product in a series of cluster integrals, yields the virial representation of the free energy: 
\begin{equation}\label{v-exp}
\begin{split}
F=F_\text{id}+F_\text{ex}&=k_\text{B}T \sum_{k=1}^M\int\rho_k(\vec r)(\ln{(\rho_k(\vec r)\Lambda^n_k)}-1)\,d\gamma \\
& + k_\text{B}TV \sum_{n=2}^\infty\, \sum_{k_1,\ldots, k_n = 1}^M \int \frac{1}{n} B_n(\Gamma_n)\,\rho_{k_1}(\vec 
r_1)\ldots \rho_{k_n}(\vec r_n)\,d\gamma_1\ldots d\gamma_n\;,
\end{split}
\end{equation}
with the ``thermal wavelength'' $\Lambda_k$ of the kinetic part $F_\text{id}$. The excess energy $F_\text{ex}$ is an 
infinite sum over virial integrals, depending on particle densities and sums over products of f-functions 
$B_n(\Gamma_n)$, with the Mayer clusters (also called diagram or graph) $\Gamma_n$ representing an unordered sum over 
all labeled, 2-connected star-diagrams $\Gamma_{n,k}$ with $n\geq 2$ nodes and counting index $k$:
\begin{equation}\label{diagram}
\Gamma_n = \sum_k \Gamma_{n,k}\;.
\end{equation}
The number of graphs is a rapidly increasing function of $n$, whose asymptotic dependence for unlabeled diagrams has 
been estimated by Riddell and Uhlenbeck to be $2^{n(n-1)/2}/n!$ \cite{uhlenbeck-ford-1,riddell}. This divergence of 
cluster integrals and the difficulties of their evaluation are the principal reasons why the virial approach is mostly 
limited to the gaseous state.

In order to go beyond the low-density limit, several alternative approaches have been developed. An early attempt has 
been taken by Reis, Frisch, and Lebowitz, resulting in the development of the scaled particle theory for hard spheres 
\cite{scaled-particle-1}. This approach is based on a heuristic but non-perturbative relation between the low- and 
high-density limit of the free energy and an analytic solution for the second virial integral 
\cite{isihara-orig,kihara-1,kihara-2}. Later on, this ansatz has been extended to convex particles based on results 
from Ishihara and Kihara, who derived $B_2$ in terms of Minkowski measures \cite{minkowski} and developed further by 
Rosenfeld into a local formulation in weight functions, suitable for density functionals \cite{rosenfeld2}. However, 
the equivalence between the volume form of the Minkowski sum of domains and their respective intersection probability 
is strictly restricted to convex surfaces and limited to two particles. It, therefore, does not generalize to higher 
order virial clusters.

Starting from the f-function of hard particles, which is a negative step-function vanishing for non-intersecting 
domains $D_i, D_j$:  
\begin{equation}\label{f-function}
f_{ij} = \begin{cases}
 -1 & \text{if}\quad D_i\cap D_j \neq 0\\
 \;\;\; 0 & \text{else} \end{cases}
\end{equation}
Rosenfeld observed that its Fourier transformed integrand factorizes into 1-particle contributions. Transforming back, 
the Mayer function can be written as the sum over a convolute of distribution and tensor valued 1-particle weight 
functions:
\begin{equation}\label{decoupling}
f_{ij}(\vec r_i-\vec r_j)= - \int C^{A_iA_j}\,w_{A_i}^i(\vec r_i-\vec r_a)w_{A_j}^j(\vec r_j-\vec r_a)\,d\gamma_a\;,
\end{equation}
with the constant and symmetric coefficient matrix $C^{A_iA_j}$ depending on the dimension of the imbedding space but 
otherwise independent of the particles' geometry. The transformation introduces the intersection coordinate $\vec r_a 
\in D_i\cap D_j$ as a new variable relative to the particle positions and orientations $\vec r_i \in D_i$. Here and in 
the following, we will omit the orientational dependence for the sake of clarity.

Rosenfeld's weight functions $w_A^i$ are the local counterparts to the Minkowski measures of integral geometry 
\cite{santalo-book,weil}. In 3 dimensions they depend on the normal vector $\vec{\hat n}$, Gaussian curvature 
$\kappa_G$, mean curvature $\bar\kappa$, curvature difference $\Delta$, surface $\sigma$, and the volume $v$:
\begin{equation}\label{weights}
\begin{split}
&w_G(\vec r_i-\vec r_a) = \frac{1}{4\pi}\kappa_G\delta(\vec{\hat n}\vec r_a) \;,\quad 
w_{\kappa L}(\vec r_i-\vec r_a) =\frac{1}{4\pi}\bar\kappa\,\vec{\hat n}^{\otimes L}\delta(\vec{\hat n}\vec r_a)\;,\\ 
&\,w_{\Delta L}(\vec r_i-\vec r_a)=\frac{1}{4\pi}\Delta\,\vec{\hat n}^{\otimes L}\delta(\vec{\hat n}\vec r_a) \;,\quad 
w_{\sigma L}(\vec r_i-\vec r_a) = \vec{\hat n}^{\otimes L} \delta(\vec{\hat n}\vec r_a) \;,\\[0.4em]
&\hspace{9em} w_v(\vec r_i-\vec r_a)= \Theta(\vec{\hat n}\vec r_a)\;,
\end{split}
\end{equation}
where the $L$-fold tensor product of the normal vector $\vec{\hat n}^{\otimes L}$ follows from a Taylor expansion of 
trigonometric functions, while the theta- and delta-functions restrict the integration to the volume, respective 
particle surface, as introduced in the appendix of \cite{korden-2}. 

Because the splitting (\ref{decoupling}) had originally been derived for spherical particles, its general dependence on 
$\Delta$ and the infinite set of tensor-valued weight functions had been obtained only later by Wertheim 
\cite{wertheim-1,wertheim-2,wertheim-3} and, independently by Mecke et~al \cite{mecke-fmt}, using the connection 
between (\ref{decoupling}) and the Gauss-Bonnet identity \cite{rosenfeld-gauss2}. Wertheim also introduced the notion 
of n-point density functions \cite{wertheim-1,wertheim-2}:
\begin{equation}\label{n-point}
n_{A_1\ldots A_n}(\vec r_{a_1},\ldots,\vec r_{a_n})=\sum_{i=1}^M \int w_{A_1}^i(\vec r_i-\vec r_{a_1})\ldots 
w_{A_n}^i(\vec r_i-\vec r_{a_n})\,\rho_i(\vec r_i)\, d\gamma_i
\end{equation}
in which any Mayer integral can be rewritten. Given the example of the third virial integral
\begin{equation}\label{third-virial}
\begin{split}
B_3 &= \frac{1}{6} \int f_{12}f_{23}f_{31}\,\rho_1\rho_2\rho_3\,d\gamma_1d\gamma_2d\gamma_3 \\
& = - \frac{1}{6} C^{A_1A_2}C^{B_2B_3}C^{C_3C_1}\int n_{A_1C_1}n_{A_2B_2}n_{B_3C_3}\,d\gamma_a d\gamma_b d\gamma_c\;,  
\end{split}
\end{equation}
Mayer clusters transform from a representation in particle positions and orientations to a corresponding representation 
in intersection coordinates $\gamma_i \to \gamma_a$. 

Thus, instead of a virial expansion in increasing powers of 1-particle densities (\ref{v-exp}), the expansion in 
n-point densities suggests an ordering by their number of intersection centers \cite{korden-2}:
\begin{equation}\label{i-exp}
F_\text{ex}=k_BTV\sum_{n=1}^\infty\int\Phi_n(\vec r_{a_1},\ldots,\vec r_{a_n})\,d\gamma_{a_1}\ldots d\gamma_{a_n}\;,
\end{equation}
whose leading order in $3$ dimensions has the generic form: 
\begin{equation}\label{first_order}
\Phi_1(\vec r_a)=-n_G\,\ln{(1-n_v)} + C_{\alpha_1\alpha_2}\frac{n_{\alpha_1}n_{\alpha_2}}{1-n_v} 
+ C_{\alpha_1\alpha_2\alpha_3}\frac{n_{\alpha_1}n_{\alpha_2}n_{\alpha_3}}{(1-n_v)^2}\;,
\end{equation}
where the volume dependence has been separated from the remaining densities $\alpha \in \{\kappa L, \Delta L, 
\sigma L\}$ and of which the Rosenfeld functional provides a first approximation \cite{tarazona-rosenfeld}:
\begin{equation}\label{ros-functional}
\begin{split}
\Phi_1^{(R)}(\vec r_a) &= -n_G\,\ln{(1-n_v)} + \frac{n_{\kappa 0}n_{\sigma 0} - n_{\kappa 1}n_{\sigma 1}}{1-n_v}\\
&\qquad\qquad + \frac{1}{24\pi}\frac{n_{\sigma 0}^3 - 3n_{\sigma 0}^{} n_{\sigma 1}^2 + \frac{9}{2} (n_{\sigma 1}^{}
n_{\sigma 2}^{} n_{\sigma 1}^{} - n_{\sigma 2}^3)}{(1-n_v)^2}\;.
\end{split}
\end{equation}
The first two parts of this polynomial in the free-volume $1-n_v$ is uniquely determined by the splitting of the second 
virial integral (\ref{decoupling}) and the scaled particle theory. The form of the third part, however, is only 
constrained by the scaling degree of the free-energy density. Several versions have therefore been proposed and tested, 
comparing its structure to analytical results and computer data \cite{roth-rev}. Its exact form will be determined in 
Sec.~\ref{sec:rose-n}.
\subsection{Resummation of Ree-Hoover Diagrams}\label{subsec:rh-diagrams}
The expansion of the free-energy functional in intersection centers (\ref{i-exp}) is not well represented by Mayer 
diagrams. 
\begin{figure}
\centering
\includegraphics[width=1.3cm,angle=-90]{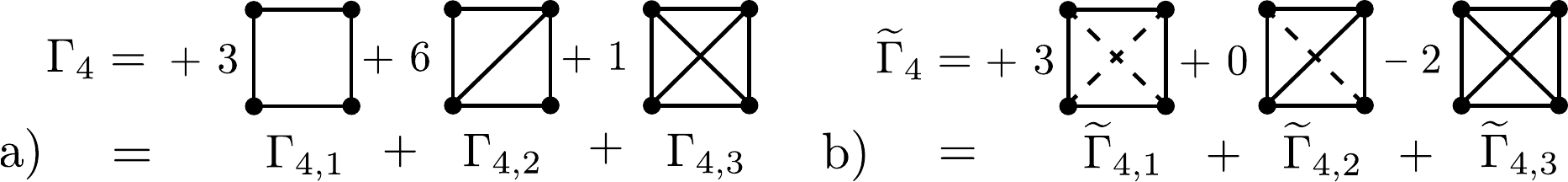}
\caption{The fourth virial diagrams in Mayer $\Gamma_4$ and Ree-Hoover $\widetilde\Gamma_4$ representation: a) The slow 
convergence of the virial expansion is partly explained by identical intersection patterns in Mayer diagrams of equal 
order. b) Ree-Hoover diagrams resolve this problem. With the additional Boltzmann functions, the allowed 
intersection of particles are constrained. Additionally, the symmetry factors of many diagrams vanish, thus reducing 
the overall number of integrals entering the virial expansion.}
\label{fig:virial}
\end{figure}
As can be seen from Fig.~\ref{fig:virial}a), any intersection network of the ring diagram $\Gamma_{4,1}$ is also found 
in $\Gamma_{4,2}$ and $\Gamma_{4,3}$, thus contributing to $\Phi_1$, $\Phi_3$, and $\Phi_4$. This redundancy, which can 
be found for any diagram of identical number of nodes, explains an observation made by Ree and Hoover in their 
numerical investigations of Mayer integrals for spheres \cite{rh-1,rh-2,rh-3}. Ordering the $n$-particle virial 
integrals by their signs into positive and negative contributions $B_n = B^+_n-B^-_n$, their individual parts are of 
comparable size $B_n^+ \simeq B_n^-$, but much larger than $B_n$ itself, often by two orders of magnitude \cite{rh-1}. 
Thus, in order to reduce the number of redundant intersection patterns, they inserted the identity $1 = e_{ij} - f_{ij}$ 
for any pair of nodes $i,j$ not bonded by an f-function. The resulting Ree-Hoover (RH) diagrams are completely 
connected graphs of f- and e-bonds, as shown in Fig.~\ref{fig:virial}b). The type of allowed intersections is therefore 
constrained and comes close to the graphical interpretation of FMT functionals as intersection networks. In the 
following, we will show that this representation not only significantly simplifies our previous derivation, but also 
allows the systematic generalization of the approximate free-energy functional to higher orders.

The previous analysis of the Rosenfeld functional started from the observation that its dependence on a single 
intersection center can only be related to diagrams which allow the particles to intersect in a common point. While 
this involves all Mayer clusters, it selects the subclass of RH-graphs without e-bonds. It therefore establishes a 
one-to-one relation between its elements of the virial series and the exclusively f-bonded RH-clusters, with their 
integration domains restricted to a single intersection center. This relation can now be used as a guideline for the 
construction of higher order functionals, summarized in four steps: 1. choose an intersection network and find the 
corresponding class of RH-diagrams, 2. determine their symmetry factors and 3. intersection probabilities, and 4. sum 
over all elements of the class. In the remaining part of the current section, we will focus on the first two subtopics, 
saving the last two items to Sec.\ref{sec:r-functional}.

For the two classes of Mayer and RH-diagrams, we introduce the following conventions:
\begin{deft}\label{def:graphs-notation}
Let $\Gamma_{n,k}$ denote a labeled, 2-path connected Mayer diagram (star-graph) of $n$ nodes and $|\Gamma_{n,k}|$ 
f-bonds. Two Mayer graphs are subgraphs $\Gamma_{n,k}\subseteq \Gamma_{n,k'}$ if they agree after removing a finite 
number of f-bonds from $\Gamma_{n,k'}$.

A node can be removed by deleting its vertex with all its bonds $\pi^{-1}: \Gamma_{n,k} \to \{\Gamma_{n-1,k'}, 
\Gamma_{n-1, t}^A\}$, resulting in a residual diagram which is either a new star-graph $\Gamma_{n-1, 
k'}$ or a linear chain with articulation points $\Gamma_{n-1, t}^A$.
\end{deft}
\begin{deft}
Let $\widetilde\Gamma_{n,k}$ be a RH-diagram with $n$ nodes and $|\widetilde\Gamma_{n,k}|$ f-bonds. 

A node, which is only linked by f-bonds, can be removed from a diagram by deleting its vertex and all its bonds, leaving 
either a new or the trivial RH-graph $\pi^{-1}: \widetilde \Gamma_{n,k} \to \{\widetilde 
\Gamma_{n-1,k'}, 0\}$. 

Two RH-graphs are subgraphs $\widetilde\Gamma_{n-m,k'} \subseteq \widetilde\Gamma_{n,k}$ if they are related by 
$\pi^{-m}=(\pi^{-1})^m$.
\end{deft}
\begin{deft}
Mayer- and RH-graphs are subgraphs, $\Gamma_{n,k}\subseteq\widetilde\Gamma_{n,k'}$, $\widetilde\Gamma_{n,k} \subseteq 
\Gamma_{n,k'}$, if they agree after removing a finite number of f-bonds and deleting all e-bonds.
\end{deft}

Because the application of $\pi^{-1}$ on a RH-graph maps to exactly one element, its inverse operation $\pi: \widetilde 
\Gamma_{n-1,k}\to\widetilde\Gamma_{n,k'}$ can be defined for $\widetilde\Gamma_{n-1,k}\neq0$, which adds a further node 
to the graph, linked by f-bonds to all $n-1$ nodes. Thus, each RH-diagram is an element of exactly one class
\begin{equation}\label{subclass}
\widetilde\Lambda_{n_0,k} = \bigcup_{m=0}^\infty \pi^m(\widetilde\Gamma_{n_0, k})\;,
\end{equation}
with the lowest subgraph $\widetilde\Gamma_{n_0,k'}\subseteq \widetilde\Gamma_{n,k}$ uniquely defined by $\pi^{-1} 
(\widetilde\Gamma_{n_0,k'})=0$. One example has already occurred in the discussion of the Rosenfeld functional. Starting 
from the single node diagram $\widetilde\Gamma_{1,1}$, each of the exclusively f-bonded diagrams is then an element of 
$\widetilde\Lambda_{1,1}$. An important property of a class of RH-diagrams is that their intersection networks can 
be chosen to coincide:
\begin{lem}\label{cor:lambda}
The intersection network of the class $\widetilde\Lambda_{n_0,k}$ is defined by its lowest subgraph.
\end{lem}
\begin{figure}
\centering
\includegraphics[width=2.4cm,angle=-90]{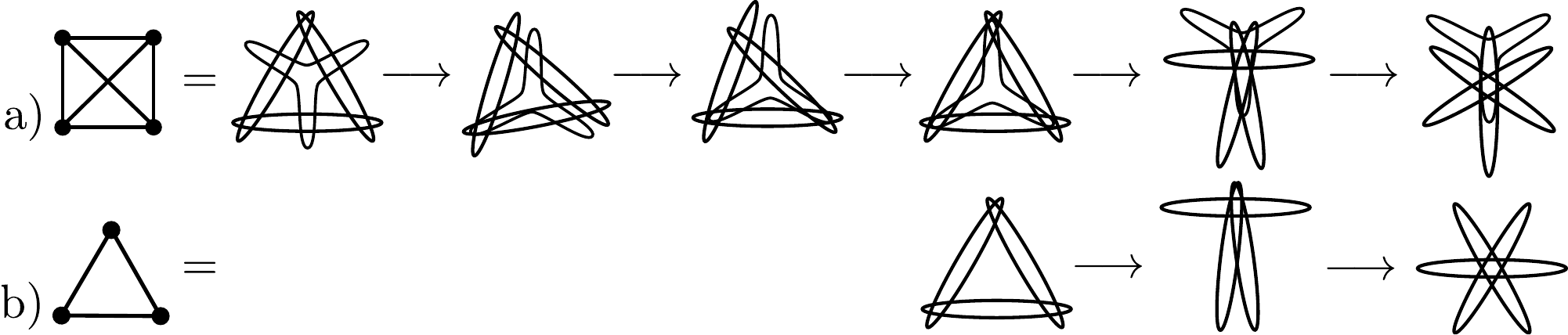}
\caption{Two RH-diagrams and their corresponding representations as intersection networks. a) The particles of the 
fully f-bonded graph $\widetilde\Gamma_{4,3}$ generically intersect in six domains, which can be successively 
contracted down to one. b) The graph $\widetilde\Gamma_{3,1}$ of the third virial cluster has three intersection 
centers, which can be shifted together into one domain.}
\label{fig:networks}
\end{figure}

This is readily seen using a graphical argument. As can be seen from Fig.~\ref{fig:virial} and Fig.~\ref{fig:networks}, 
for a given diagram $\widetilde\Gamma_{n,k}$ there are two types of nodes, those exclusively linked by f-bonds and 
those belonging to the lowest subdiagram, or ``backbone'', $\widetilde\Gamma_{n_0,k'}$. Choose a specific intersection 
network for this subgraph. Any further particle can then be attached to this network by overlapping it with all of the 
previous intersection centers, thus keeping the number of centers unchanged. 

Focusing again on the Rosenfeld functional, the first element of the class $\widetilde\Lambda_{1,1}$ with a non-trivial 
intersection center is $\widetilde\Gamma_{2,1}$. Thus all RH-graphs with the second virial diagram as its subgraph can 
be reduced to an intersecting network whose particles overlap in exactly one center. However, one might also choose 
$\widetilde\Gamma_{3,1}$ as the leading element. Then, the sum over all RH-integrals of the class $\widetilde 
\Lambda_{3,1}$, contracted to three intersection centers, yields $\Phi_3$. Thus, once a lowest subgraph and its 
intersection pattern has been selected, its functional is uniquely fixed by its corresponding class. This solves the 
first of the previously stated four problems.

As the representation of the functional has been shifted from Mayer- to RH-diagrams, it is also necessary to rewrite 
the functional itself (\ref{v-exp}) in this new basis. This transformation has already been shown by Ree and Hoover to 
be a linear combination, weighted by the sign of f-numbers \cite{rh-3,leroux}:
\begin{equation}
\widetilde\Gamma_{n,k} = \sum_{k'}^{\widetilde\Gamma_{n,k}\subseteq \Gamma_{n,k'}} \Gamma_{n,k'}\;\;,\quad
\Gamma_{n,k} = \sum_{k'}^{\Gamma_{n,k}\subseteq \widetilde\Gamma_{n,k'}} 
(-1)^{|\Gamma_{n,k}|-|\widetilde\Gamma_{n,k'}|}\, \widetilde\Gamma_{n,k'}\;.
\end{equation}
When inserted into the sum of Mayer clusters of $n$ vertices (\ref{diagram}), the summation order can be exchanged
\begin{equation}\label{two-gamma-reps}
\Gamma_n = \sum_{k'}\Gamma_{n,k'} = \sum_{k'}\sum_{k}^{\Gamma_{n,k'}\subseteq \widetilde\Gamma_{n,k}} 
(-1)^{|\Gamma_{n,k'}| - 
|\widetilde\Gamma_{n,k}|}\,\widetilde\Gamma_{n,k} = \sum_{k} a_{n,k}\,\widetilde\Gamma_{n,k}\;,
\end{equation}
leading to the ``star-content'' of a RH-graph \cite{rh-3}:
\begin{equation}\label{star-content}
a_{n,k} = \sum_{k'}^{\Gamma_{n,k'}\subseteq \widetilde\Gamma_{n,k}} (-1)^{|\Gamma_{n,k'}|-|\widetilde\Gamma_{n,k}|}\;.
\end{equation}

An important property of the star-content, proven by Ree and Hoover, is its recursion relation under removal of 
an exclusively f-bonded vertex point \cite{rh-3}. When the operator $\pi^{-1}$ from Def.~\ref{def:graphs-notation} is 
applied to $\Gamma_n$, the sum separates into star-diagrams $\Gamma_{n-1,k'}$ and 1-path connected graphs with 
articulation points $\Gamma_{n-1, t}^A$, each weighted by their sign of f-numbers in order to avoid overcounting of 
contributions:
\begin{equation}\label{rh-sum}
\begin{split}
\pi^{-1}(\Gamma_n)&=\sum_k\sum_{k'}^{\Gamma_{n-1,k'}\subset\Gamma_{n,k}}(-1)^{|\Gamma_{n,k}|-|\Gamma_{n-1,k'}|}\,
\Gamma_{n-1,k'}\\
&\hspace{6em} +\sum_k\sum_{t}^{\Gamma_{n-1,t}^A\subset\Gamma_{n,k}}(-1)^{|\Gamma_{n,k}|-|\Gamma_{n-1,t}^A|}\, 
\Gamma_{n-1,t}^A\;.
\end{split}
\end{equation}

Focusing on the first part, the order of summation can be exchanged, resulting in a sum over those 
$\Gamma_{n,k}$-graphs, which lead to the same $\Gamma_{n-1,k'}$ diagram after removing a node. If this node is linked 
by $m$ bonds, these bonds can be distributed in $(n-1)!/(m!(n-m)!)$ ways among the residual $n-1$ vertex points, each 
weighted by a sign-factor of $(-1)^m$. The first sum therefore simplifies to:
\begin{equation}\label{rh-non-zero}
\begin{split}
& \sum_{k'} \sum_k^{\Gamma_{n,k}\supset \Gamma_{n-1,k'}} (-1)^{|\Gamma_{n,k}|-|\Gamma_{n-1,k'}|}\, \Gamma_{n-1,k'}
= \sum_{k'} \sum_{m=2}^{n-1}\binom{n-1}{m}(-1)^m \, \Gamma_{n-1, k'}\\
& = (n-2)\,\Gamma_{n-1}\;.
\end{split}
\end{equation}

The second part is a sum over $\Gamma_{n,k}$ and its 1-path connected subgraphs $\Gamma_{n-1,t}^A$. Again, the order of 
summation can be exchanged, resulting in a weighted sum over all permutations of $0\leq m\leq n-1$ bonds linked to the 
residual $n-1$ nodes. In order to perform this sum, observe that a 1-path connected diagram is a linear chain of 
star-graphs, connected by articulation points, here indicated by $'*'$:
\begin{equation}
\Gamma_{n-1,t}^A = \Gamma_{n_1,k_1}*\Gamma_{n_2,k_2}*\ldots*\Gamma_{n_p,k_p}\;,
\end{equation}
with $n-1=n_1+n_2+\ldots+n_p$ nodes. Any partition of $m$ bonds between the vertex $P_n$, which is to be removed from 
$\widetilde \Gamma_{n,k}$, and the non-articulation points can now be compensated by a copy of this same diagram with 
an additional bond between $P_n$ and one of the articulation points. Having the same subgraph but differing by one 
f-bond, the two diagrams cancel each other in Eq.~(\ref{rh-sum}). The sum therefore decouples into the partition of 
bonds on non-articulation vertices and articulation ones. The flanking graphs $\Gamma_{n_1, k_1}$, $\Gamma_{n_p,k_p}$ 
are each joined by one articulation point, while the center ones carry two. Their individual contributions to the 
weighted sum are therefore 
\begin{equation}
\sum_{m=0}^{n_i-2}\binom{n_i-2}{m}(-1)^m=0\;,\qquad \sum_{m=1}^{n_i-1}\binom{n_i-1}{m}(-1)^m=-1
\end{equation}
yielding an overall factor of $(-1)^2 = 1$ and leaving a sum over all permutations of bonds between the articulation 
points and $P_n$:
\begin{equation}\label{rh-zero}
\sum_t \sum_k^{\Gamma_{n,k}\supset \Gamma_{n-1,t}} (-1)^{|\Gamma_{n,k}|-|\Gamma_{n-1,t}|}\, \Gamma_{n-1,t}
= \sum_k \sum_{m=0}^{p-1}\binom{p-1}{m}(-1)^m \, \Gamma_{n-1,t} = 0\;.
\end{equation}
Combining the two results (\ref{rh-non-zero}), (\ref{rh-zero}), the total sum of (\ref{rh-sum}) reduces to the 
recursion relation of Ree and Hoover \cite{rh-3}:
\begin{equation}
\pi^{-1}(\Gamma_n) = (n-2)\,\Gamma_{n-1}\;.
\end{equation}

An analogous relation can be derived for RH-diagrams, inserting the recursion relation into (\ref{two-gamma-reps}) and 
observing that $\pi^{-1}$ removes $n-1$ f-bonds from a RH-diagram $\widetilde\Gamma_{n,k}$:
\begin{equation}
\begin{split}
\pi^{-1}(\Gamma_n) &= \sum_k \pi^{-1}(a_{n,k}\,\widetilde\Gamma_{n,k})
= (-1)^{n-1} \sum_k a_{n,k}\,\widetilde\Gamma_{n-1,k}\\
&= (n-2)\,\Gamma_{n-1} = (n-2)\sum_{k'} a_{n-1,k'}\,\widetilde\Gamma_{n-1,k'}\;.
\end{split}
\end{equation}
Comparing terms, the resulting recursion relation is again independent of $k$:
\begin{equation}
a_{n,k} = (-1)^{n-1}(n-2)\,a_{n-1,k'}
\end{equation}
and by repeated application yields the star-content \cite{rh-3}:
\begin{equation}\label{star-content-rec}
a_{n,k} = (-1)^{\binom{n}{2}-\binom{m}{2}}\frac{(n-2)!}{(m-2)!}\, a_{m,k'}\;.
\end{equation}
Thus, once it has been calculated for the lowest diagram of a class $\widetilde\Lambda_{n_0,k}$, the coefficients for 
all further graphs are known. 

So far, we have focused on labeled graphs, ignoring that the physical particles in the partition function are 
indistinguishable. Transferring to unlabeled RH-diagrams therefore introduces an additional sum over all inequivalent 
permutations of the labels on the vertex points. For its derivation the e-bonds can be ignored, as they do not change 
the graph's symmetry, yielding the ``symmetry factor'' $\sigma_{n,k}$ already known from Mayer clusters. In combination 
with the star-content, we define the combinatorial prefactor of the virial clusters in RH-graphs:
\begin{equation}\label{rh-sym-fact}
\widetilde\sigma_{n,k}=-\frac{1}{n!}\sigma_{n,k}\,a_{n,k}\;\,,
\end{equation}
where the additional factor $1/n!$ has been inserted for later convenience. The last step in proving the resummability 
of cluster integrals for a given class $\widetilde\Lambda_{n_0,k}$ therefore requires the derivation of a recursion 
relation for $\sigma_{n,k}$.

The symmetry factor counts the number of inequivalent labelings of a RH- or Mayer graph $\Gamma_{n,k}$ under the
permutation group $S_n$. Its group elements operate on the product of f-functions, which define a representation space 
by $\lambda:\Gamma_{n,k}\to\text{prod}(f_{ij})$ and are symmetric under exchange of positions and indices:
\begin{equation}
\gamma:=\{f_{ij}=f_{ji},\;f_{ij}f_{kl}=f_{kl}f_{ij}\}\;,\;\;\gamma\,\lambda(\Gamma_{n,k})=\lambda(\Gamma_{n,k})\;.
\end{equation}
The group of equivalent labelings is then the automorphism group of its diagram and a subset of the symmetric group 
$S_n$
\begin{equation}
\text{Aut}(\Gamma_{n,k}) = \{g\in S_n\;|\; g\lambda(\Gamma_{n,k}) = \gamma\,\lambda(\Gamma_{n,k}) \}
\end{equation}
subject to the invariance relation $\gamma^{-1}\circ g = \text{id}$. Correspondingly, the set of inequivalent labelings 
is generated by the coset $S_n/\text{Aut}(\Gamma_{n,k})$, whose number of elements is the symmetry factor of the graph
\begin{equation}\label{coset-dim}
\sigma_{n,k} = \frac{|S_n|}{|\text{Aut}(\Gamma_{n,k})|}\;.
\end{equation}
An extensive list of Mayer diagrams and their automorphism groups has been tabulated by Uhlenbeck and Riddell 
\cite{uhlenbeck-ford-1,riddell}.

In generally, determining the automorphism group of a given graph is non-trivial. However, if it is known for the 
lowest element of a class $\widetilde\Lambda_{n_0,k}$, the groups of all further diagrams are known:
\begin{lem}
Let $\widetilde\Gamma_{n,k}\in\widetilde\Lambda_{n_0,k}$ with lowest element $\widetilde\Gamma_{n_0,k}$. Its 
automorphism group then factorizes into the direct product:
\begin{equation}\label{aut}
\text{Aut}(\widetilde\Gamma_{n,k}) = S_{n-n_0}\times \text{Aut}(\widetilde\Gamma_{n_0,k})\;.
\end{equation}
\end{lem}
The proof is as follows: Because $\widetilde\Gamma_{n_0,k}$ is the lowest element of the class, $\pi^{-1} \widetilde 
\Gamma_{n_0,k}=0$, each vertex is connected to at least one e-bond, whereas the remaining $n-n_0$ nodes are only linked 
by f-bonds. Shifting an f-bonded label to a vertex with an e-bond would therefore result in an inequivalently labeled 
diagram. Thus, f-bonded nodes can only be permuted under themselves by $S_{n-n_0}$, completing the proof.

With the recursion relation for the star content (\ref{star-content-rec}) and the dimension of the automorphism group
\begin{equation}
|\text{Aut}(\widetilde\Gamma_{n,k})| = (n-n_0)!\;|\text{Aut}(\widetilde\Gamma_{n_0,k})|\;,
\end{equation}
the RH-diagrams of a given class have the symmetry factor:
\begin{lem}\label{cor:sym-factor}
Let $\widetilde\Gamma_{n,k}$ be an element of the class $\widetilde\Lambda_{n_0,k}$ of RH-graphs with lowest element 
$\widetilde\Gamma_{n_0,k}$. Its symmetry factor is:
\begin{equation}\label{sig-tilde}
\begin{split}
\widetilde\sigma_{n,k} & = -(-1)^{\binom{n}{2}-\binom{n_0}{2}}\binom{n-2}{n_0-2} 
\frac{a_{n_0,k}}{|\text{Aut}(\widetilde\Gamma_{n_0,k})|}\;\;:\qquad n_0 \geq 3\\
\widetilde\sigma_{n,k} & =\;\;\;\, (-1)^{\binom{n}{2}}\frac{1}{n(n-1)}\hspace{7.6em}:\qquad  n_0=1
\end{split}
\end{equation}
\end{lem}
The case $n_0\geq 3$ follows from the previous calculations. But $n_0=1$ and $n_0=2$ have to be considered separately, 
because the recursion relation (\ref{star-content-rec}) only applies to $n_0\geq 2$, while (\ref{aut}) requires 
$\widetilde\Gamma_{n_0,k}$ to be the lowest element of the class. Observing that $\widetilde\Gamma_{1,1} =\pi^{-1} 
(\widetilde\Gamma_{2,1})$ and $a_{2,1}=a_{1,1}=1$ closes the proof.

It is instructive to derive the symmetry factors for the diagrams of Fig.~\ref{fig:virial}. $\widetilde\Gamma_{4,3}$ 
is a fully f-bonded graph and therefore belongs to $\widetilde\Lambda_{1,1}$, already covered by (\ref{sig-tilde}). The 
next diagram $\widetilde\Gamma_{4,2}$ has star-content $a_{4,2}(|)=0$. Thus, all graphs with a single e-bond drop out 
of the virial series $\widetilde\sigma_n(|) = 0$. Actually, the same applies to roughly half of the RH-diagrams, 
improving the convergence of the virial expansion. The final example $\widetilde\Gamma_{4,1}$ has two separate e-bonds, 
star-content $a_{4,1}(||)=1$, and cyclic permutation symmetry $\text{Aut}(\widetilde\Gamma_{4,1}) = \mathbb{Z}_4 \times 
\mathbb{Z}_2$. The prefactors of all related graphs 
are therefore
\begin{equation}
\widetilde\sigma_n(||) = (-1)^{\binom{n}{2}}\frac{1}{16}(n-2)(n-3)\;.
\end{equation}

Eq.~(\ref{sig-tilde}) solves the second of the four problems stated at the beginning of this section and also holds 
the solution to the fourth one. As has been shown before in 3 dimensions \cite{korden-3}, the intersection probability 
of a network of overlapping particles decouples into a convolute of curvature forms $K_n(\vec r_a)$ for each 
intersection center with $n$ particles. This observation is sufficient to write down a generic functional for a given 
intersection network and RH-class:
\begin{thm}\label{prop:1}
Let $\widetilde\Gamma_{n_0,k}$ be the lowest element of the class $\widetilde\Lambda_{n_0,k}$. The free-energy 
functional density of a network with $p\geq 1$ intersection centers is then determined by
\begin{equation}\label{inter-form-1}
\Phi_1(\widetilde\Gamma_{1,1}| \vec r_a) = \sum_{n\geq 2} \frac{1}{n(n-1)} \int K_n(\vec r_a, \{\gamma_i\})
\rho(\gamma_1)\ldots \rho(\gamma_n)\, d\gamma_1\ldots d\gamma_n
\end{equation}
for $n_0=1$, while higher order functionals for $n_0 > 3$, are derived by 
\begin{equation}\label{inter-form-2}
\begin{split}
&\Phi_p(\widetilde\Gamma_{n_0,k}| \vec r_{a_1}, \ldots, \vec r_{a_p}) = -(-1)^{\binom{n_0}{2}} 
\frac{a_{n_0,k}}{|\text{Aut}(\widetilde\Gamma_{n_0,k})|} \sum_{n\geq n_0} \binom{n-2}{n_0-2}\\
&\qquad\quad \times \int  K_{n_1}(\vec r_{a_1}, \{\gamma_{i_1}\})\ldots K_{n_p}(\vec r_{a_p}, 
\{\gamma_{i_p}\}) \rho(\gamma_1)\ldots \rho(\gamma_n)\, d\gamma_1\ldots d\gamma_n
\end{split}
\end{equation}
with the integral kernel $K_{n_j}(\vec r_a,\{\gamma_i\})$ determining the intersection probability of $n_j$ particles 
which intersect in $\vec r_a$ and with particle positions $\gamma_i\,$ for $i=1,\ldots, n_j$.
\end{thm}
The first half of the proof determines the numerical prefactor, combining the symmetry factor of (\ref{sig-tilde}) and 
the sign of f-bonds $(-1)^{|\widetilde\Gamma_{n,k}|}$, which is not accounted for by the intersection forms. It 
partially 
cancels with the sign of $\widetilde\sigma_{n,k}$ using
\begin{equation}
|\widetilde\Gamma_{n,k}|= |\widetilde\Gamma_{n_0,k}| + \binom{n}{2}-\binom{n_0}{2}\;,
\end{equation}
leaving a constant only depending on the lowest element of $\widetilde\Lambda_{n_0,k}$ and an $n$-dependent binomial 
coefficient. The second half of the proof deals with the factorization of the intersection probability into integral 
kernels, which will be the topic of the next section. There it will also be shown, how to generalize the functionals to 
e-bonds and mixtures of particles.
\section{Intersection Probability in N Dimensions}\label{sec:r-functional}
In this section we complete the proof of Theorem~\ref{prop:1}, determining the intersection probability for networks of 
overlapping particles. First, the basic ideas of integral geometry are summarized, determining the intersection 
probability of clusters of particles with one intersection center and then generalized to networks of clusters.

Integral geometry determines the intersection probability between manifolds under translations and rotations with 
respect to their imbedding space. It has a long history \cite{santalo-book,weil}, but its modern representation has 
been founded by Blaschke and Santalo, and further generalized by Chern to $n$ dimensions using differential forms 
\cite{chern-1,chern-2,chern-3}. Chern also explained the connection between the kinematic measure and the Euler form, 
determining the Blaschke-Santalo-Chern equation for two intersecting manifolds \cite{chern-1,santalo-book}. Up to a 
sign, this result coincides with Rosenfeld's decoupling of the second virial integral (\ref{decoupling}). Using this 
approach, we calculated the intersection probability for an arbitrary set of particles overlapping in a common domain, 
rederiving the Rosenfeld functional for $3$ dimensions. This calculation will be simplified in the following and 
extended to arbitrary dimension $n$.

For now, let $\Sigma_k$ be a set of identical, $n-1$ dimensional, smooth, orientable, compact, boundary free, 
Riemannian manifolds with particle index $k=1,\ldots,N$, imbedded into $D_k:\Sigma_k \hookrightarrow \mathbb{R}^n$. For 
physical reasons we will further assume that the particles $D_k$ only have cavities that are accessible to all other 
particles. Each domain is then covered by coordinate patches with an orthonormal, positively oriented coordinate frame 
$(\hat e_1^{(k)}, \ldots, \hat e_n^{(k)})$ at each point $p\in D_k$, with the normal vector $\hat e_n^{(k)}$ 
pointing into the outside direction of the surface $\Sigma_k = \partial D_k$. Being a differential manifold, the 
geometry of $\Sigma$ and $D$ are represented by vielbein and connection forms
\begin{equation}
\theta_i = \hat e_i dp\;,\quad \omega_{ij}\, = \hat e_i d\hat e_j\quad \text{for}\quad i,j=1,\ldots, n\;,
\end{equation}
and the intrinsic forms of torsion and curvature
\begin{equation}
T_i = d\theta_i - \omega_{ij}\wedge \theta_j\;,\quad \Omega_{ij} = d\omega_{ij} - \omega_{ik}\wedge\omega_{kj}\;,
\end{equation}
with the torsion $T_i = 0$ vanishing for a Riemannian manifold. Here and in the following we implicitly assume a sum 
over pairs of identical indices. Coordinate frames of different patches are connected by $g_{ij}\in\text{SO}(n)$, 
transforming vectors $e_i' = g_{ij}\,e_j$ and differential forms
\begin{equation}\label{trans-form}
\theta_i' = g_{ij}\theta_j\;,\quad \omega_{ij}' = g_{ik}^{-1}\omega_{kl}g_{lj}+g_{ik}^{-1}dg_{kj}\;,\quad 
\Omega_{ij}' = g_{ik}^{-1}\Omega_{kl}g_{lj}\;.
\end{equation}

Each Riemannian manifold also defines a semisimple group \cite{helgason}, which for the imbedding space $\mathbb{R}^n$ 
is the isometric or Euclidean group $\text{ISO}(n) = \text{SO}(n) \ltimes \mathbb{R}^n$. Its Lie algebra 
$\text{iso}(n)$ has a representation in the vielbein and connection forms generating translations and rotations
\begin{equation}
\begin{pmatrix}
\omega_{ij} & \theta_i \\
-\theta_j & 0
\end{pmatrix}
\;\;\in\;\; \text{iso}(n)
\end{equation}
and whose volume form, or Haar measure, is the ``kinematic measure''
\begin{equation}\label{iso-measure}
K(\mathbb{R}^n) =  \bigwedge_{1\leq i< j\leq n} \omega_{ij}\bigwedge_{1\leq i\leq n} \theta_i\;.
\end{equation}
An alternative interpretation uses the representation as a chain of cosets
\begin{equation}
\text{SO}(k)/\text{SO}(k-1) = S^{k-1}\;,\quad \text{vol}(S^{k-1}) = 2\pi^{k/2}/\Gamma(k/2)\;,
\end{equation}
which allows to rewrite products of connection forms into integral measures of spheres
\begin{equation}
dS^{k-1} = \omega_{1,k}\wedge\ldots\wedge \omega_{k-1,k}\;.
\end{equation}

With a particle $D$ imbedded, the vector space of $\mathbb{R}^n$ splits into the tangential $T\Sigma$ and normal space 
$N\Sigma$ along with a similar decoupling of the rotation group into $\text{SO}(n-1)$ and the coset $\text{SO}(n)/ 
\text{SO}(n-1)$. The kinematic measure (\ref{iso-measure}) separates correspondingly into the three parts
\begin{equation}\label{red-measure}
K(\partial D)=\bigwedge_{\alpha=1}^{n-1}\omega_{\alpha\, n}\bigwedge_{i=1}^n 
\theta_i\,\wedge\,d\text{vol}(\text{SO}(n-1))
\end{equation}
of the curvature (also Euler or highest Chern-Simons) form, the measure of translations, and the volume form of 
$\text{SO}(n-1)$. 

This result readily generalizes to two and more overlapping domains which intersect in at least a common point. 
Introducing the notation for identical particles
\begin{equation}
\Sigma^{k_1}\cap D^{k_2} := \underbrace{\Sigma\cap \ldots\cap \Sigma}_{k_1\text{-fold}}\cap 
\underbrace{D\cap\ldots\cap D}_{k_2\text{-fold}}\;,
\end{equation}
the intersection $D^m$ is again a $n$ dimensional subset of $\mathbb{R}^n$ and therefore applies to the reduced form 
of the kinematic measure (\ref{red-measure}). When inserted, $K(\partial D^m)$, the boundary of the intersection domain 
can be expanded in the series
\begin{equation}\label{inter-sum}
\partial D^m = \sum_{1\leq k\leq m}^{k\leq n} \binom{m}{k} \Sigma^k \cap D^{m-k}\;,
\end{equation}
setting all terms $\Sigma^k$ of $k\geq n+1$ to zero. This restriction of the sum is a useful consequence of the 
following rules of $K$:
\begin{lem}
Let $D$ be a domain in $\mathbb{R}^n$ with hypersurface $\Sigma$. The kinematic measure $K$ of intersecting manifolds 
then has the properties:
\begin{align}
K(a_1\Sigma_1+a_2\Sigma_2) & = a_1 K(\Sigma_1)+a_2 K(\Sigma_2)\label{cor-a}\\[0.2em]
K(\Sigma^{k_1}\cap D^{k_2})& = K(\Sigma^{k_1})\wedge K(D)^{k_2}\label{cor-b}\\[0.2em] 
K(\Sigma_1\cap \Sigma_2)   & = K(\Sigma_2\cap \Sigma_1)\label{cor-c}\\[0.2em]
K(\Sigma^k)                & = 0\quad \text{for}\;\; k \geq n+1\label{cor-d}
\end{align}
for $\Sigma_1\cap \Sigma_2=0$ and the constants $a_1,a_2\in \mathbb{N}$:
\end{lem}
The proof is as follows: By construction, the differential form (\ref{iso-measure}) is a locally defined probability 
measure, which decouples for $\Sigma_1\cap \Sigma_2=0$, thus showing (\ref{cor-a}). The second identity follows from 
the observation that the kinematic measure of a point $\text{pt}\in\mathbb{R}^n$ moving in the domain $D$ is trivial 
$K(\text{pt}\cap D)=K(D)$. Setting $\text{pt}\in \Sigma^{k_1}\cap D^{k_2-1}$ then proves (\ref{cor-b}) by iteration. 
While the local measure and set theoretic relation $\Sigma_1 \cap \Sigma_2=\Sigma_2\cap \Sigma_1$ explains 
(\ref{cor-c}). The last equation (\ref{cor-d}), which justifies the finite sum (\ref{inter-sum}), implies that the 
intersection probability between a point $\text{pt} \in \Sigma^n$ and a hypersurface $\Sigma$ is of measure zero. This 
follows from $K$ being defined on the tangential space of $\Sigma^k$, spanned by the normal and tangential vectors 
$(\hat e_1, \ldots, \hat e_{n-k}, \hat e_n^{(1)}, \ldots, \hat e_n^{(k)})$, proving that the coordinate basis is not 
defined for $k\geq n+1$ and therefore vanishes for dimensional reasons.

Applying these rules to the boundary of intersections (\ref{inter-sum}), yields the sum
\begin{equation}\label{measure-gen}
K(\partial D^m)=\sum_{1\leq k\leq m}^{k\leq n} \binom{m}{k} K(\Sigma^k)\wedge K(D)^{m-k}\;,
\end{equation}
with $K(D)$ and the curvature forms $K(\Sigma^k)$ decoupled. For $m=1$ this reproduces (\ref{red-measure}). But to 
derive the remaining differential forms we need an explicit representation of the orthonormal and positively oriented 
vector field at $\Sigma^k$. For simplicity, let us define a fixed ordering of surfaces $\Sigma_1\cap \ldots \cap 
\Sigma_k$. The vector 
\begin{equation}\label{vector}
(\hat e_1, \ldots, \hat e_{n-k}, \hat e_n^{(1)}, \ldots, \hat e_n^{(k)}) \in TD^k
\end{equation}
then defines a coordinate frame at the intersection, spanned by the $k$ outward pointing normal vectors and the $n-k$ 
tangential directions which can be chosen by (\ref{trans-form}) to coincide with all surfaces. However, this frame is 
neither orthonormal nor positively orientated.

The Gram-Schmidt transformation turns this vector frame into an orthonormal basis
\begin{equation}\label{trafo1}
v_A = \begin{cases} 
B_{ab} e_n^{(b)} & \text{for}\;\; a,b = 1,\ldots, k\\
\quad\, e_\alpha  & \text{for}\;\; \alpha = 1,\ldots, n-k
\end{cases}
\end{equation}
whose upper triangular matrix $B_{ab}\in \text{Gl}(k, \mathbb{R})$ combines into the general coordinate mapping:
\begin{equation}
B_{AB} = \begin{pmatrix} B_{ab} & 0 \\ 0 & 1_{\alpha\beta} 
\end{pmatrix}\;.
\end{equation}
The coordinate frames of individual surfaces are now related by a rotation of the normal vectors
\begin{equation}\label{trafo2}
\eta_A = \begin{cases} 
G_{ab} v_b & \text{for}\;\; a,b = 1,\ldots, k\\
\quad\, v_\alpha  & \text{for}\;\; \alpha = 1,\ldots, n-k
\end{cases}
\end{equation}
generated by $G_{ab}\in \text{SO}(k, \mathbb{R})$ and extended to the matrix representation
\begin{equation}
G_{AB} = \begin{pmatrix} G_{ab} & 0 \\ 0 & 1_{\alpha\beta} 
\end{pmatrix}\;.
\end{equation}
Combining both transformations, the connection forms $\eta_{AB} = \eta_A d\eta_B$ at an intersection point can be 
written in the basis of the tangential $\omega_{\alpha \beta}$ and normal directions $\omega_{\beta n}^{(a)}$ of the 
coordinate frames of $\Sigma_a$ keeping the normal vector $de_n^{(a)} = 0$ constant 
\begin{equation}
\eta_{AB} = \begin{cases} 
\eta_{ab}=G^{-1}_{ac}\,dG_{cb}\\
\eta_{a\beta} = G_{ab}\,B_{bc}\,\omega_{\beta n}^{(c)}\\
\eta_{\alpha\beta} = \omega_{\alpha\beta}
\end{cases}\;.
\end{equation}

The curvature form, which generalizes (\ref{red-measure}) to the intersection space $\Sigma^k$, is the product of 
connection forms with one axial direction fixed. Let us choose $\eta_k$ as the new normal vector, leaving $G_{ak}$ to 
be an element of the coset space $\text{SO}(k)/ \text{SO}(k-1)=S^{k-1}$ 
\begin{equation}\label{euler-form}
\begin{split}
\bigwedge_{\beta=1}^{n-k}\eta_{\beta k} \bigwedge_{a=1}^{k-1}\eta_{ak} & = (\bigwedge_{\beta=1}^{n-k} 
G_{ka}B_{ab}\,\omega_{\beta n}^{(b)}) \, (\bigwedge_{a=1}^{k-1} G^{-1}_{ab}\,dG_{bk})\\
& = \text{Pf}^{\;n-k}_{1,\ldots, k}(GB\omega)\wedge d\text{vol}_G(S^{k-1})\;,
\end{split}
\end{equation}
introducing the notation $\text{Pf}^{\;n-k}_{1,\ldots,k}$ for the Pfaffian of an $n-k$-form of the ordered set 
of surfaces $\Sigma_1 \cap\ldots\cap\Sigma_k$.

In the derivation of (\ref{euler-form}) we ignored the orientation of the coordinate frame of $T\Sigma^k$ which is 
not fixed by the Gram-Schmidt process. However, changing the orientation, e.g. by permuting the order of the particles 
in (\ref{trafo1}), changes the sign of (\ref{euler-form}) by $(-1)^{n-1}$. For $n$ odd, this poses no problem. For $n$ 
even, however, the orientation of the vector frame is relevant and has to be chosen positively orientated to ensure a 
positive surface form. This uniquely defines the intersection form $K_m$ introduces in Theorem~\ref{prop:1}:
\begin{thm}\label{prop:2}
Let $D_{i_1}\cap \ldots\cap D_{i_m}$ be a set of $n$-dimensional domains overlapping in the common intersection point 
$\vec r_a \in D^m$ and assigned with a positively orientated coordinate frame. Denote by $|S^{k-1}|$ the volume of the 
$S^{k-1}$ sphere and by $\Theta(D)$ the step function confining the integral measure to the volume of $D$. Then, the 
integral kernel of the functionals (\ref{inter-form-1}) and (\ref{inter-form-2}) has the form:
\begin{equation}\label{kernel}
\begin{split}
K_{i_1\ldots i_m}(\vec r_a,\{\gamma_i\})& =\frac{1}{|S^{n-1}|}\sum_{1\leq k\leq m}^{k\leq n} \binom{m}{k} 
\int_{G\in S^{\,k-1}} \text{Pf}(GB\omega)_{i_1\ldots i_k}^{n-k} \wedge d\text{vol}_G(S^{k-1}) \\
& \hspace{13em} \times \Theta(D_{i_{k+1}})\ldots \Theta(D_{i_m})\;,
\end{split}
\end{equation}
with an integration over the inner area of the Euler sphere $S^{k-1}$.
\end{thm}

This result follows from inserting the curvature form (\ref{euler-form}) into the kinematic measure (\ref{measure-gen}) 
and observing that the vectors $\vec r_a$ are already determined by the coordinates $\{\gamma_i\}$ of the 1-particle 
densities. The integration over $\gamma_i$ therefore shifts part of the kinematic measure into the definition of the 
functionals (\ref{inter-form-1}) and (\ref{inter-form-2}) and leaves the integral kernel with the step functions 
$\Theta(D)$ and the differential form $\text{Pf}(GB\omega)^{n-k}$. Finally, the normalization $|S^{n-1}|$ has been 
added to compensate the volume form introduced by (\ref{euler-form}). 

The curvature form (\ref{kernel}) defines the integral kernel for one intersection center. To generalize this result to 
arbitrary networks of overlapping particles, let us introduce the following notation:
\begin{deft}
Networks of intersecting particle domains are represented by products of $\gamma_a^{i_1\ldots i_m}$ for each 
intersection center $\vec r_a \in D_{i_1}\cap \ldots \cap D_{i_m}$ of the intersecting domains $D_{i_1},\ldots, 
D_{i_m}$.
\end{deft}
When applied to the third virial cluster, shown in Fig.~\ref{fig:networks}b), the generic diagram $\Gamma_{3,1}$ is an 
intersection network of pairwise overlapping domains, which can be successively contracted from three centers into
networks of two and one 
\begin{equation}
\gamma_a^{i_1i_2}\gamma_b^{i_2i_3}\gamma_c^{i_1i_3}\to\gamma_a^{i_1i_2i_3}\gamma_c^{i_1i_3}\to\gamma_a^{i_1i_2i_3}\;.
\end{equation}
While $\Gamma_{4,3}$ has the generic network of six centers, with consecutive contraction into five to one intersection 
domains, as shown in Fig.~\ref{fig:networks}a)
\begin{equation}
\begin{split}
&\gamma_a^{i_1i_2}\gamma_b^{i_2i_3}\gamma_c^{i_3i_1}\gamma_d^{i_1i_4}\gamma_e^{i_2i_4}\gamma_f^{i_3i_4}
\to \gamma_a^{i_1i_2i_4}\gamma_b^{i_2i_3}\gamma_c^{i_3i_1}\gamma_e^{i_2i_4}\gamma_f^{i_3i_4}
\to \gamma_a^{i_1i_2i_4}\gamma_b^{i_2i_3i_4}\gamma_c^{i_3i_1}\gamma_f^{i_3i_4}\\
&\quad \to \gamma_a^{i_1i_2i_4}\gamma_b^{i_2i_3i_4}\gamma_c^{i_1i_3i_4}
\to \gamma_a^{i_1i_2i_3i_4}\gamma_c^{i_1i_3i_4}
\to \gamma_a^{i_1i_2i_3i_4}\;.
\end{split}
\end{equation}
This notation for intersection networks readily translates to the integral kernel:
\begin{lem}\label{cor:net}
Let $\gamma_a^i$ and $\gamma_b^j$ denote intersection clusters at centers $\vec r_a$ and $\vec r_b$ with particle 
indices $i=\{i_1,\ldots, i_p\}$ and $j=\{j_1,\ldots, j_q\}$. And let $e_{k_1k_2}$ denote a Boltzmann function with 
arbitrary particle indices $k_1,k_2$. Introducing the notation $K(\gamma_a^{i_1\ldots i_m}) = K_{i_1\ldots i_m}(\vec 
r_a)$ for the intersection form $K$, the integral kernel for intersection networks is a linear and multiplicative 
operator
\begin{align}
K(\gamma_a^i + \gamma_b^j)& = K_i(\vec r_a) + K_j(\vec r_b)\label{net_a}\\
K(\gamma_a^i\gamma_b^j)& = K_i(\vec r_a) K_j(\vec r_b)\label{net_b}\\
K(\gamma_a^i e_{k_1k_2}) &=  K_i(\vec r_a)\, e_{k_1k_2}\;.\label{net_c}
\end{align}
\end{lem}
The first identity (\ref{net_a}) is a generalization of (\ref{cor-a}), while the product structure (\ref{net_b}) is a 
consequence of the properties of f-bonds, whose intersection centers overlap independently. Boltzmann functions, 
however, do not contribute to the curvature form, but define constraints on the integration domain, justifying 
(\ref{net_c}).

Less trivial, however, is the meaning of the indices. In $\gamma_a^{i_1\ldots i_p}$, each index $i_k\in i$ points to an 
individual domain $D_{i_k}$ of the $i_k=1,\ldots, N$ particles, while the same index in $K_{i_1\ldots i_p}$ points to 
the compound $i_k=1, \ldots, M$ represented by the 1-particle density $\rho_{i_k}$. This change of meaning is a 
consequence of the thermodynamic limit introduced in Sec.~\ref{subsec:intersects} and corresponds to the substitution 
\begin{equation}
D_k \to \sum_{i_k=1}^M D_{i_k} \rho_{i_k}\;.
\end{equation}
When inserted into the kinematic measure (\ref{measure-gen}), the combinatorial prefactors remain unchanged, while the 
differential forms are weighted by the density functions. Thus each particle index has to be paired by a corresponding 
density function:
\begin{equation}
\sum_{i_1,i_2,i_3=1}^M K(\gamma_a^{i_1i_2}\gamma_b^{i_2i_3}\gamma_c^{i_3i_1})\rho_{i_1}\rho_{i_2}\rho_{i_3}\;.
\end{equation}

The generalization to mixtures completes the proof of Theorem~\ref{prop:1}. And together with the integral kernel 
(\ref{kernel}) and its algebraic structure, we have the necessary tools to systematically derive the density functional 
for any given class of intersection diagrams. However, to do so efficiently, let us introduce some further notation. 
First observe that for any $k$, the curvature form (\ref{euler-form}) can be written as a volume form
\begin{equation}\label{weights-form}
\begin{split}
\text{Pf}(GB\omega)^{n-k}_{i_1\ldots i_k} & = \det{(GBh)}_{i_1\ldots i_k} \bigwedge_{\alpha=1}^{n-k}\theta_\alpha\\ 
& = \int_{\mathbb{R}^n} [J\,\det{(GBh)}]_{i_1\ldots i_k} \delta(\Sigma_{i_1})\ldots \delta(\Sigma_{i_k})
\bigwedge_{i=1}^n \theta_i\;,
\end{split}
\end{equation}
introducing the curvature matrix $\omega_{\alpha,n} = h_{\alpha\beta}\theta_\beta$ and the delta-function 
$\delta(\Sigma)$, which projects the integration domain from $\mathbb{R}^n$ to the tangential space of the surface 
$\Sigma$. Correspondingly, products of delta-functions project to the intersection space $\Sigma_{i_1}\cap \ldots\cap 
\Sigma_{i_k}$ associated with the Jacobi determinant
\begin{equation}
J_{i_1\ldots i_k}= \det{(\hat e_1, \ldots, \hat e_{n-k},\hat e_n^{(i_1)}, \ldots, \hat e_n^{(i_k)})}\;.
\end{equation}

This transformation allows to factor out the overall volume form and to rewrite the integral kernel (\ref{kernel}) as 
the derivative of a generating function:
\begin{lem}\label{cor:weights}
Define the weight functions for $k=1, \ldots, n$ intersecting surfaces
\begin{equation}\label{weights-2}
w^{i_1\ldots i_k}_k = \frac{1}{|S^{n-1}|}\int_{G\in S^{k-1}} [J\,\det{(GBh)}]_{i_1\ldots i_k} d\text{vol}_G (S^{k-1}) 
\delta(\Sigma_{i_1})\ldots \delta(\Sigma_{i_k})
\end{equation}
and the weight function for the domain
\begin{equation}
w_0^{i_1}=\Theta(D_{i_1})\;.
\end{equation}
The integral kernel (\ref{kernel}) of the intersection form is then the functional derivative of the product of 
$w_0$-weights:
\begin{equation}\label{cor-gen}
K(\gamma^{i_1\ldots i_m}_a) = \mathcal{D}_a\; w_0^{i_1}(\vec r_{ai_1})\ldots w_0^{i_m}(\vec r_{ai_m})\;,
\end{equation}
with the derivative on the weight functions at intersection $\vec r_{ai}=\vec r_a - \vec r_i$ defined by
\begin{equation}\label{deriv}
\mathcal{D}_a = \sum_{k=1}^n\sum_{i_1\ldots i_k}\frac{1}{k!} w^{i_1\ldots i_k}_k(\vec r_{ai_1},\ldots \vec r_{ai_k}) 
\frac{\delta^k}{\delta w_0^{i_1}(\vec r_{ai_1})\ldots \delta w_0^{i_k}(\vec r_{ai_k})}\;,
\end{equation}
operating on two and more $w_0$-weight functions.
\end{lem}
The proof follows by differentiating the integral and significantly simplifies by introducing the extended 
Kronecker-delta and summation rule:
\begin{equation}
\begin{split}
&\delta^j_i := \frac{\delta}{\delta w_0^i(\vec r_{ai})} w_0^j(\vec r_{aj})\;\;,\qquad
\sum_j \delta^j_i w_0^j(\vec r_{aj}) = w_0^i(\vec r_{ai})\;,\\
&\hspace{4em} \frac{\delta}{\delta w_0^i(\vec r_{ai})} w_0^j(\vec r_{bj}) = 0 \quad \text{for}\quad 
\vec r_a\neq \vec r_b\;.
\end{split}
\end{equation}
Less trivial is the observation that the curvature form (\ref{weights-form}) and therefore the weight functions are 
symmetric in the particle indices. This is a consequence of the coupled coordinate system of $T\Sigma^k$, which can be 
diagonalized using the $\text{SO}(n-k) \times \text{SO}(k)$ invariance of its vector basis. Its explicit calculation 
gives no further insight and therefore is postponed to App.~\ref{sec:appendix_a}.

The derivative (\ref{deriv}), which is a consequence of (\ref{inter-sum}), simplifies the construction of a functional 
in two aspects: first, it hides the sum over the $k=1,\ldots, n$ differential forms in the intersection kernel 
(\ref{kernel}). Second, it leaves the volume weight $w_0^i$ as the only variable in the sums of (\ref{inter-form-1}), 
(\ref{inter-form-2}). Thus, each functional is the derivative of a generating function in the virial series of $w_0^i$. 
In combination with Theorem~\ref{prop:1}, this reveals that:
\begin{lem}\label{lem:con-rad} 
The virial series of the free-energy functional for a finite number of intersection centers is convergent.
\end{lem}
The proof is as follows: It is sufficient to show that the generating function is a convergent series. If $\gamma_0$ is 
the intersection diagram of the lowest element $\widetilde\Gamma_{n_0,k}$ with $p$ intersection centers. Then, the 
integral kernel of any higher order element $\pi^m\widetilde\Gamma_{n_0,k}$ decouples:
\begin{equation}
\int K(\pi^m \gamma_0)\rho^{n_0+m} = (x_{a_1\ldots a_p})^m \int K(\gamma_0)\rho^{n_0}\;,
\end{equation}
with the density function defined by
\begin{equation}\label{x-def}
x_{a_1 \ldots a_p} = \int w_0^{i}(\vec r_{a_1i})\ldots w_0^{i}(\vec r_{a_pi})\rho_{i}d\gamma_i\;.
\end{equation}
When inserted into the functionals (\ref{inter-form-1}), (\ref{inter-form-2}), the sum simplifies to a polynomial in 
$x$ with singularities in $x=1$ and $x=0$. However, the virial series is finite in the low-density limit and therefore  
$x=0$ a regular point. This leaves $x=1$ as the only singularity. At thermodynamic equilibrium and with all 
intersection centers collapsed into a single point, the packing fraction approaches $\eta=1$, proving that the 
functional is convergent for $\eta<1$.

This final result proves the resummability of the approximate density functional and solves the last of the four 
problems presented at the beginning of Sec.~\ref{subsec:rh-diagrams}.
\section{The Rosenfeld Functional and Beyond}\label{sec:rose-n}
This last section presents four examples, which provide some insight into the construction of functionals. We begin  
with the integral kernel for $3$ dimensions and derive the functional with one intersection center. It is then shown 
that Rosenfeld's result is an approximation for almost perpendicular normal vectors. The remaining examples then focus 
on the functionals with two, three, and four intersection centers.

Starting point for the construction of any functional is the derivation of the integral kernel. This has to be done 
once for any dimension and is independent of the particle geometry or boundary conditions. For $n=3$ dimensions, there 
exists three differential forms:

The case $k=1$ is elementary. With $G=B=1$ and assuming a positively oriented orthonormal coordinate frame, the 
Pfaffian reduces to the Euler form $\text{Pf}^{\,2}_{i_1}(GB\omega) = \omega_{13}^{(i_1)} \wedge \omega_{23}^{(i_1)}$ 
and a trivial integral over the $S^0$-sphere
\begin{equation}\label{diff_1}
\int \text{Pf}^{\,2}_{i_1}(GB\omega)\wedge d\text{vol}(S^0)=\omega_{13}^{(i_1)}\wedge\omega_{23}^{(i_1)}\;.
\end{equation}

For $k=2$, the Gram-Schmidt scheme and $\text{SO}(2)$ rotation have the matrix representation:
\begin{equation}
\begin{split}
B = \begin{pmatrix}
1 & 0 & 0 \\ \displaystyle{-\frac{\cos{(\phi_{i_1i_2})}}{\sin{(\phi_{i_1i_2})}}} & 
\displaystyle{\frac{1}{\sin{(\phi_{i_1i_2})}} }& 
0 \\ 0 & 0 & 1
\end{pmatrix}
\;&,\quad 
G = \begin{pmatrix}
\;\;\;\cos{(\alpha)} & \sin{(\alpha)} & 0 \\ -\sin{(\alpha)} & \cos{(\alpha)} & 0 \\ 0 & 0 & 1        
\end{pmatrix}\\
\phi_{i_1i_2} = \arccos{(\hat e_3^{(i_1)}\hat e_3^{(i_2)})}\quad\;\; &,\qquad\quad\;  
0\leq \alpha \leq \phi_{i_1i_2}\leq 2\pi\;,
\end{split}
\end{equation}
depending on the intersection angle $\phi_{i_1i_2}$. The associate Pfaffian yields the 1-form:
\begin{equation}
\text{Pf}^{\,1}_{i_1i_2}(GB\omega) = \bigl[\cos{(\alpha)}-\frac{\cos{(\phi_{i_1i_2})}}{\sin{(\phi_{i_1i_2})}} 
\sin{(\alpha)}\bigr] \omega_{13}^{(i_1)} 
+ \frac{1}{\sin{(\phi_{i_1i_2})}}\sin{(\alpha)}\,\omega_{13}^{(i_2)}
\end{equation}
whose integral over $\alpha\in S^1$ simplifies to
\begin{equation}
\int_{\alpha\in S^1}\text{Pf}^{\,1}_{i_1i_2}(GB\omega)\wedge d\text{vol}_\alpha(S^1) 
= \frac{1-\cos{(\phi_{i_1i_2})}}{\sin{(\phi_{i_1i_2})}}\bigr[ \omega_{13}^{(i_1)} + \omega_{13}^{(i_2)} \bigr]\;.
\end{equation}

Finally, for $k=3$ the integral over the curvature form (\ref{euler-form}) reduces to the area of the spherical 
triangle $\Delta\subseteq S^2$
\begin{equation}
\int_{G\in S^2}\text{Pf}^{\,0}_{i_1i_2i_3}(GB\omega)\wedge d\text{vol}_G(S^2) 
= \text{area}(\Delta \subseteq S^2) = 2\pi - \phi_{i_2i_3}^{i_1} - \phi_{i_1i_3}^{i_2} - \phi_{i_1i_2}^{i_3}
\end{equation}
whose value is determined by the Euler or dihedral angles $\phi_{i_1i_2}^{i_3}$ \cite{hsiung}:
\begin{equation}\label{euler-angle}
\phi^{i_1}_{i_2i_3} = \arccos{(\hat E_{i_2}\hat E_{i_3})}\;,\qquad 
\hat E_{i_1} = \frac{\hat e_3^{(i_2)}\times \hat e_3^{(i_3)}}{|\hat e_3^{(i_2)}\times \hat e_3^{(i_3)}|}\;.
\end{equation}
Comparing these results to the definition (\ref{weights-2}) and taking into account the permutation symmetry in the 
particle indices, we obtain the three weight functions:  
\begin{equation}\label{muli-weight}
\begin{split}
w_1^{i_1}=\frac{1}{4\pi}&\det{(h^{(i_1)})} \delta(\Sigma_{i_1}),\;
w_2^{i_1i_2}=\frac{1}{2\pi}\frac{1-\cos{(\phi_{i_1i_2})}}{\sin{(\phi_{i_1i_2})}} 
[Jh_{12}]^{(i_1)}\delta(\Sigma_{i_1})\delta(\Sigma_{i_2})\\[0.4em]
&\qquad\; w^{i_1i_2i_3}_3 = \frac{1}{4\pi}(2\pi-3\phi_{i_2i_3}^{i_1})J_{i_1i_2i_3} \delta(\Sigma_{i_1}) 
\delta(\Sigma_{i_2}) \delta(\Sigma_{i_3})\;.
\end{split}
\end{equation}
These agree with the weight function $w_2^{i_1i_2}$ first derived by Wertheim \cite{wertheim-1,wertheim-2} and 
rediscovered in \cite{mecke-fmt}, while $w_3^{i_1i_2i_3}$ first occurred in \cite{marechal-1}.

Next, we will illustrate the application of the derivative and generating function on the functional restricted to one 
intersection center $\Phi|_1$. It is therefore of type (\ref{inter-form-1}), and the virial series sums up to the 
generating function in the $x$-variable (\ref{x-def}) of the volume weight:
\begin{equation}\label{gen-1}
\begin{split}
\Phi|_1(\widetilde\Gamma_{1,1}, \vec r_a)& =\int \sum_{m=2}\frac{1}{m(m-1)}K(\gamma_a^{i_1\ldots i_m})\rho_{i_1}\ldots 
\rho_{i_m}\,d\gamma_{i_1}\ldots d\gamma_{i_m}\\
&=\mathcal{D}_a\int \sum_{m=2} \frac{1}{m(m-1)} w_0^{i_1}(\vec r_a)\ldots w_0^{i_m}(\vec r_a)\rho_{i_1}\ldots 
\rho_{i_m}\,d\gamma_{i_1}\ldots 
d\gamma_{i_m} \\
&=\mathcal{D}_a\sum_{m=2}\frac{1}{m(m-1)}x_a^m = \mathcal{D}_a[(1-x_a)\ln{(1-x_a)}+x_a]\;.
\end{split}
\end{equation}
Introducing the weight density
\begin{equation}
n_k(\vec r_a) = \int w_k^{i_1\ldots i_k}(\vec r_{ai_1}, \ldots, \vec r_{ai_k})\,\rho_{i_1}(\vec r_{i_1})\ldots 
\rho_{i_k}(\vec r_{i_k})\,d\gamma_{i_1}\ldots d\gamma_{i_k}\;,
\end{equation}
the derivative of the generating function yields the three terms
\begin{equation}\label{1-cent}
\Phi|_1(\widetilde\Gamma_{1,1}, \vec r_a) =-n_1\ln{(1-n_0)}+\frac{1}{2}\frac{n_2}{1-n_0}+\frac{1}{6}
\frac{n_3}{(1-n_0)^2}
\end{equation}
whose structure is well known from the Rosenfeld functional (\ref{ros-functional}). It is well established that the 
weight functions (\ref{weights}) derive from a Taylor expansion in the $sin$ and $cos$ terms of the intersection 
angles. This has been shown before for $n_2$ in \cite{mecke-fmt,mecke-fmt2} and by using differential forms 
\cite{korden-2}. For $n_3$, however, we made an inconvenient choice of coordinates for $\Delta\in S^2$ and give here a 
clearer argument. A more detailed discussion that also includes the zero-dimensional limit will be presented in 
\cite{m_k_m}.

Let us introduce the notation $\hat e_i := \hat e_n^{(i)}$ for the $i=1,2,3$ normal vectors of $\Sigma_1$, $\Sigma_2$, 
$\Sigma_3$ and $e_{ij}:=\hat e_i \hat e_j$ for their scalar product. Further assume that the three vectors 
approximately set up an orthonormal basis
\begin{equation}
\hat e_1 \approx \hat e_2\times \hat e_3\;.
\end{equation}
The area of the spherical triangle is then $\Delta \approx \pi/2$ with the Jacobi determinant $J_{123}= \det{(\hat e_1, 
\hat e_2,\hat e_3))} \approx 1$ and Euler angles $\phi_{ij}^k$ close to zero. Thus, expanding the argument of $arccos$ 
up to 5'th order
\begin{equation}
z=\hat E_1\hat E_2=(e_{23}e_{13}-e_{12})(1+\frac{1}{2}e_{23}^2)(1+\frac{1}{2}e_{13}^3) + \mathcal{O}(e_{ij}^5)\;,
\end{equation}
yields the small-angle correction of $\Delta$:
\begin{equation}
\begin{split}
2\pi-3\arccos{(z)} &= \frac{\pi}{2} + 3(z+ \frac{1}{6}z^3) + \mathcal{O}(z^5)\\
&=\frac{\pi}{2}-3\bigl[e_{12}-e_{23}e_{13}+\frac{1}{2}e_{12}(e_{23}^2+e_{13}^2)+\frac{1}{6}e_{12}^3\bigr]
+\mathcal{O}(e_{ij}^4)
\end{split}
\end{equation}
where only $e_{12}^3$ is from order $z^3$. Rewritten in the weight densities (\ref{weights}), $n_3$ is to first order 
in the Euler angle:
\begin{equation}
n_3 = \frac{\pi}{2}n_{\sigma 0}^3-3n_{\sigma 0}n_{\sigma 1}^2+3[n_{\sigma 1}n_{\sigma 2}n_{\sigma 1} 
-n_{\sigma 1} n_{\sigma 2} n_{\sigma 3}]-\frac{1}{2}n_{\sigma 3}^2 n_{\sigma 0} + \mathcal{O}(e_{ij}^4\rho^3).
\end{equation}
The second term agrees with the Rosenfeld functional, while the prefactor of the third is $4.5$ compared to our $3$, 
which is acceptable given that the terms of (\ref{ros-functional}) have been fitted to numerical data. The fourth term 
disagrees, but is of lower order. More remarkably, the prefactor of the first term disagrees also. Choosing a different 
point of reference for the Taylor expansion can set the coefficient of the leading term from $\pi/2$ to $0$, but never 
to $1$. Here, Tarazona's term for the 3-particle intersection is a better approximation \cite{tarazona-rosenfeld}, 
which also includes additional terms from the third order in $z$. A more detailed discussion can be found in 
\cite{m_k_m}.

Our next example derives the functional for three intersection centers, whose lowest RH-diagram is $\widetilde 
\Gamma_{3,1}$, which belongs to the class $\widetilde\Lambda_{1,1}$. Its intersection diagram is the exact third virial 
cluster $\gamma_a^{i_1i_2}\gamma_b^{i_2i_3}\gamma_c^{i_3i_1}$ show in Fig.~\ref{fig:networks}b). The next particle 
added to the triangle diagram has to overlap with all three intersection centers, as illustrated by the fourth figure 
of Fig.~\ref{fig:networks}a) and represented by $\gamma_a^{i_1i_2i_4} \gamma_b^{i_2i_3i_4} \gamma_c^{i_3i_1i_4}$. 
Adding more particles includes copies of the last one, corresponding to the intersection pattern
\begin{equation}\label{3-cent-g}
\gamma_a^{i_1i_2i_4\ldots i_m}\gamma_b^{i_2i_3i_4\ldots i_m}\gamma_c^{i_3i_1i_4\ldots i_m}\;.
\end{equation}
To further simplify the notation, let us define $w_a^i := w_0^i(\vec r_{ai})$ for the volume weight at intersection 
center $\vec r_a$ and particle position $\vec r_i$. The functional (\ref{inter-form-1}) for the 3-center diagrams then 
factorizes into weight densities of 2 and 3 centers:
\begin{equation}
\begin{split}
&\sum_{m=3}^\infty \frac{1}{m(m-1)}  \int K(\gamma_a^{i_1i_2i_4\ldots i_m}\gamma_b^{i_2i_3i_4\ldots 
i_m}\gamma_c^{i_1i_3i_4\ldots i_m})\,\rho_{i_1}\ldots \rho_{i_m}\,d\gamma_{i_1}\ldots d\gamma_{i_m}\\
&=\mathcal{D}_a \mathcal{D}_b \mathcal{D}_c \sum_{m=3}^\infty \frac{1}{m(m-1)}\int (w_a^{i_1}w_c^{i_1}\rho_{i_1}) 
(w_a^{i_2}w_b^{i_2}\rho_{i_2}) (w_b^{i_3}w_c^{i_3}\rho_{i_3})\\
&\hspace{7em} \times (w_a^{i_4}w_b^{i_4}w_c^{i_4}\rho_{i_4})\ldots (w_a^{i_m}w_b^{i_m}w_c^{i_m}\rho_{i_m})\,
d\gamma_{i_1}\ldots d\gamma_{i_m}\;.
\end{split}
\end{equation}
Introducing the $x$-densities of (\ref{x-def}) and adding the second virial, we finally arrive at the functional 
restricted to three and less intersection centers:
\begin{equation}\label{3-cent}
\begin{split}
&\Phi|_3(\widetilde\Gamma_{1,1}, \vec r_a, \vec r_b, \vec r_c) = \frac{1}{2}\mathcal{D}_a x_a^2\\
&\qquad\quad  + \mathcal{D}_a \mathcal{D}_b \mathcal{D}_c \;\Bigl[ \frac{x_{ab} x_{bc} 
x_{ac}}{x^3_{abc}}\bigl((1-x_{abc})\ln{(1-x_{abc})} + x_{abc} 
-\frac{1}{2}x^2_{abc}\bigr)\Bigr]\;.
\end{split}
\end{equation}

With three intersection centers, it is exact in the second and third virial order, but also significantly more complex 
due to additional correlation and autocorrelation functions such as $w_1^{i_4}(\vec r_a) w_1^{i_4}(\vec r_b) 
w_1^{i_4}(\vec r_c)$, which have been shown by Wertheim to be nontrivial to evaluate 
\cite{wertheim-2,wertheim-3,wertheim-4}.

Nevertheless, Lemma~\ref{cor:weights} is an efficient tool to derive new functionals. To compare different levels of 
approximation, let us summarize all cases of up to four intersection centers. With one center covered by (\ref{gen-1}), 
the next level is the 2-center approximation, beginning with the partially contracted third virial cluster, shown in  
Fig.~\ref{fig:networks}b). All further intersection diagrams are then of the form:
\begin{equation}
\gamma_a^{i_1i_2i_3\;i_4\ldots i_m}\gamma_b^{i_2i_3\; i_4\ldots i_m}\;.
\end{equation}
The case of three centers has already been covered by (\ref{3-cent-g}). This leaves the functional with four 
intersection centers of the two RH-classes $\widetilde\Gamma_{4,1}$ and $\widetilde\Gamma_{4,3}$, while the 
star-content of $\widetilde\Gamma_{4,2}$ vanishes. The corresponding intersection graphs are:
\begin{equation}
\begin{split}
\gamma_a^{i_1i_2\;i_5\ldots i_m}\gamma_b^{i_2i_3\;i_5\ldots i_m}&\gamma_c^{i_3i_4\;i_5\ldots i_m} 
\gamma_d^{i_1i_4\;i_5\ldots i_m}e_{i_1i_3}e_{i_2i_4}\\
&+\; \gamma_a^{i_1i_2}\gamma_b^{i_2i_3\;i_4\ldots i_m} \gamma_c^{i_1i_3\;i_4\ldots i_m} \gamma_d^{i_2i_4\ldots i_m}\;,
\end{split}
\end{equation}
with the Boltzmann functions $e_{ij}$ included. Determining their generating functions and adding the lower virial 
terms, yields the complete list of functionals up to four intersections
\begin{align}
\Phi|_1 &= \mathcal{D}_a[(1-x_a)\ln{(1-x_a)}+x_a]\nonumber\\
\Phi|_2 &= \frac{1}{2}\mathcal{D}_a x_a^2 + \frac{1}{2}\mathcal{D}_a\mathcal{D}_b \frac{x_a+x_b}{x_{ab}}
\Bigl[(1-x_{ab})\ln{(1-x_{ab})} + x_{ab} - \frac{1}{2}x_{ab}^2\Bigr] \nonumber\\
\Phi|_3 &= \frac{1}{2}\mathcal{D}_a x_a^2 + \mathcal{D}_a \mathcal{D}_b \mathcal{D}_c \;\frac{x_{ab} x_{bc} 
x_{ac}}{x^3_{abc}}\Bigl[(1-x_{abc})\ln{(1-x_{abc})} + x_{abc} -\frac{1}{2}x^2_{abc}\Bigr]\nonumber\\
\Phi|_4 &= \frac{1}{2}\mathcal{D}_a x_a^2 + \frac{1}{6}\mathcal{D}_a \mathcal{D}_b \mathcal{D}_c 
x_{ab}x_{bc}x_{ac}\label{final-list}\\
&+  \mathcal{D}_a \mathcal{D}_b \mathcal{D}_c \mathcal{D}_d \frac{x_{ac}x_{bc}x_{abd}}{x_{bcd}^4}
\Bigl[(1-x_{bcd})\ln{(1-x_{bcd})} + x_{bcd} - \frac{1}{2}x^2_{bcd} - \frac{1}{6} x_{bcd}^3\Bigr]\nonumber\\
& + \frac{1}{8} \mathcal{D}_a \mathcal{D}_b \mathcal{D}_c \mathcal{D}_d 
\frac{y_{ab|cd}y_{ad|bc}}{(1-x_{abcd})^3}\;,\nonumber
\end{align}
where we replaced $x_a\to (x_a+x_b)/2$ in $\Phi|_2$ to satisfy the diagram's symmetry and introduced the definition of 
the Boltzmann weighted density
\begin{equation}
y_{ab|cd} = \int (w_a^i w_b^i\rho_i)(w_c^j w_d^j\rho_j)\, e_{ij}d\gamma_i d\gamma_j\;.
\end{equation}

The formalism of Lemma~\ref{cor:weights} is not restricted to the free energy but applies to any thermodynamic object 
representable by Mayer or RH-graphs. As a final example, let us consider the leading order of the pair-correlation 
function $g_2$. As has been shown for spheres in \cite{rh-3,rh-4}, the dominating RH-graphs are again the fully 
f-bonded diagrams with two root points connected by an e-bond. The network with the lowest number of intersection 
centers is therefore
\begin{equation}
\gamma_a^{i_1\;i_3\ldots i_m}\gamma_b^{i_2\;i_3\ldots i_m}\,e_{i_1i_2}\;,
\end{equation}
whose combinatorial prefactors have been shown in \cite{rh-4} to combine to $1$. The sum over all diagrams then yields 
$g_2$ approximated by two intersection centers
\begin{equation}
g_{i_1i_2}|_2(\vec r_a,\vec r_b, \vec r_{i_1},\vec r_{i_2}) = \mathcal{D}_a \mathcal{D}_b 
\frac{e_{i_1i_2}w_a^{i_1}w_b^{i_2}}{1-x_{ab}}\;,
\end{equation}
without a sum over the particle indices $i_1,i_2$.

The derivation of these examples is a result of simple algebra. But the increasing order of n-point densities gives a 
first impression of the difficulties one will have to overcome to minimize such functionals. The two centers of 
$\Phi|_2$ define a rotation axis over which the weight functions can individually be averaged without further 
constraints. Significantly more difficulties arise from $\Phi|_3$. The three centers form a ring, which cause the 
positions and orientations of the particles to couple. Still, the integrals can be evaluated by Wertheim's 
application of the Radon transformation \cite{wertheim-2}. Finally, $\Phi|_4$ contains the elements of RH-class 
$\widetilde \Lambda_{4,1}$, introducing e-bonds and therefore constraints on the particles not to overlap. This adds a 
further level of complexity to the model, which effectively limits the DFT approach to these four functionals.
\section{Discussion and Conclusion}\label{sec:conclusion}
The last examples have shown that the construction of a functional for hard particles is a rather simple task. For a 
given virial order, one first chooses the intersection network and derives its generating function in the volume 
density. In a second step, one derives the weight functions from the intersection kernel and finally performs the 
derivatives. But the non-locality of the problem has not vanished; instead it is hidden in the convolute of the 
intersection coordinates. 

The rapidly increasing complexity with the number of intersection centers effectively imposes an upper bound on the 
order of the functional to be applied. On the other hand, the long range ordering of the particles in the high density 
limit also defines a lower bound on the number of intersection centers. This is because the minimization process cannot 
produce a particle density with a lower symmetry as the invariance group of the functional. The underlying intersection 
network therefore has to be complex enough to reduce the isometric group of the imbedding space to a sufficiently small 
subgroup. But here one can be optimistic. The Rosenfeld functional with the highest rotational and translational 
symmetry already predicts, e.g., the nematic phase. And the next leading order $\Phi|_2$ only depends on 1- and 2-point 
densities without further geometric constraints due to e-bonds or rings which, e.g., complicates the evaluation of 
$\Phi|_4$ and $\Phi|_3$.

But it might not be even necessary to go beyond the leading order. As the last example has already indicated, any 
correlation function can be approximated by intersection diagrams and represented in weight densities. This will be 
investigated in a forthcoming paper and applied to couple soft interactions to the hard-particle reference fluid. The 
soft potential, however, significantly deforms the hard-particle phase structure, so that a detailed knowledge of the 
hard-particle phase becomes less important.

For the derivation of the functional we made rather strong assumptions about the particles' geometry, demanding their 
surfaces to be smooth and bounded. Both requirements can be weakened. Actually, it is sufficient for the manifolds to be 
compact, twice differentiable, and Riemannian or representable by triangulation. And instead of the Euclidean imbedding 
space it is possible to use complex, projective or hyperbolic spaces, as long as they have infinite volume. Finite 
volumes can be constructed from them by adding external potentials. These spaces and imbeddings are probably of less 
physical interest. But it is known from computer simulations that the phase structure of a system depends sensitively on 
the geometry. Thus, instead of intrinsic objects, such as Chern or Chern-Simons classes, it is possible to characterize 
manifolds by their many-particle properties. The connection between single- and many-particle properties is the 
integral kernel, whose derivation closely parallels that of Chern-Simons forms. The virial coefficients of the 
free-energy expansion therefore contain the integral values of these classes, either obtained by computer simulations, 
minimizing a functional, or other methods. This allows to identify the numerical value of any virial order with its 
respective differential forms. Manifolds can then be characterized by these numbers and tested by mixing them with 
simpler particles such as spheres of different sizes. 

The explicitly known dependence of the non-local functional therefore might provide a direct view into the 
thermodynamics of hard particles that is only rivalled by scalar field theory and the Ising model.
\begin{acknowledgements}
The author wishes to thank Andr\'{e} Bardow and Kai Leonhard for support of this work as part of his PhD thesis. He 
also gratefully acknowledges Matthieu Marechal for many helpful discussions and sharing his insight into Ree-Hoover 
diagrams, also Annett Schwarz and Christian Jens for improving the manuscript.

\noindent This work was performed as part of the Cluster of Excellence "Tailor-Made Fuels from Biomass", funded by the 
Excellence Initiative of the German federal and state governments.
\end{acknowledgements}
\appendix
\section{Appendix}\label{sec:appendix_a}
The current section completes the proof Lemma~\ref{cor:weights} that the weight functions (\ref{weights-2}) are 
symmetric in the particles indices and permuted by $\text{SO}(k)$ rotations.

For the surfaces $\Sigma_1, \ldots, \Sigma_k$, let us define the $k$ coordinate frames
\begin{equation}\label{coord-frame}
\Sigma_m\;:\;(\underbrace{\hat e_1^{(m)}, \ldots, \hat e_{n-m}^{(m)}}_{\text{tangential}}, \underbrace{\hat 
e_{n-m+1}^{(m)}, \ldots, \hat e_n^{(m)}}_{\text{normal}})\quad\text{for}\quad m=1,\ldots, k
\end{equation}
and the Pfaffian form (\ref{weights-form})
\begin{equation}
\text{Pf}(GB\omega)^{n-k}_{1\ldots k} = \det{(GBh)}_{1 \ldots k}\bigwedge_{\alpha=1}^{n-k}\theta_\alpha^{(1)}\;.
\end{equation}
This choice of the vector base defines a specific ordering of the particles' indices $1,\ldots,k$, which can be viewed 
as a nested intersection of surfaces
\begin{equation}\label{nested}
(((\Sigma_1 \cap \Sigma_2) \cap \Sigma_3) \cap \Sigma_4) \ldots\;, 
\end{equation}
with $\Sigma_1$ as the reference system. It therefore fails the permutation invariance of the intersection probability. 
However, not the curvature form but the kinematic measure (\ref{cor-c}) has to be $\text{SO}(n-k)\times \text{SO}(k)$ 
invariant. We therefore have to prove the permutation symmetry of the integral
\begin{equation}\label{kin-meas-2}
\begin{split}
&\int K(\Sigma_1\cap\ldots\cap \Sigma_k)\wedge K(D_2)\wedge \ldots\wedge K(D_k)\\
&\qquad =\int \det(GBh)_{1\ldots k}\bigwedge_{\alpha=1}^{n-k}\theta^{(1)}_\alpha 
\frac{d\text{vol}(\text{SO}(n-k))}{|\text{SO}(n-k)|}\wedge 
K(D_2)\wedge\ldots\wedge K(D_k)\;.
\end{split}
\end{equation}

Generalizing Chern's arguments for two intersection surfaces \cite{chern-2,santalo-book}, the translational measure of 
$\Sigma^k$ has been extended by $\text{SO}(n-k)$ to the kinematic measure of $\mathbb{R}^{n-k}$. This leaves us to 
rotate the normal vectors $\hat e_n^{(2)}, \ldots, \hat e_n^{(k)}$ into the tangential direction of $\Sigma_1$. For the 
nested coordinate frame (\ref{nested}), we can chose the $\text{SO}(m)$ transformations of the ``normal'' directions to 
rotate $\hat e_n^{(m)}$ into the basis of $\Sigma_1$ while keeping the vectors of $\Sigma_2,\ldots, \Sigma_{m-1}$ fixed:
\begin{equation}
\Sigma_m \to \Sigma_1\;:\quad \text{SO}(m)/\text{SO}(m-1) = S^{m-1}\;.
\end{equation}
Introducing the spherical vector $(x_1^{(m)}, \ldots, x_m^{(m)}) \in S^{m-1}$, the normal vector of $\Sigma_m$ 
transforms into the basis of $\Sigma_1$
\begin{equation}
\hat e_n^{(m)} = \sum_{p=1}^m\;x_p^{(m)}\;\hat e_{n-p+1}^{(1)}\;\quad \text{for}
\quad \sum_{p=1}^m (x^{(m)}_p)^2 = 1
\end{equation}
and their dual basis forms
\begin{equation}
\begin{split}
\theta_n^{(m)}& = \sum_{p=2}^m\; x_p^{(m)}\, \theta_{n-p+1}^{(1)} = x_m^{(m)}\theta_{n-m+1}^{(1)} + 
\mathcal{O}(\theta_\alpha^{(1)})\\
&\hspace{7em}\text{for}\quad n-m+2 \leq \alpha\leq n-1\;.
\end{split}
\end{equation}
Here, the symbol $\mathcal{O}(\theta_\alpha^{(1)})$ indicates elements, which drop out when inserted into the skew 
symmetric product (\ref{kin-meas-2}) as they already occur at a lower $m$. This transforms the normal directions into 
the tangential forms of $\Sigma_1$
\begin{equation}
\theta_n^{(2)}\wedge\ldots \wedge \theta_n^{(k)} = x_2^{(2)}\ldots x_k^{(k)}\, \theta_{n-1}^{(1)}\wedge \ldots\wedge 
\theta_{n-k+1}^{(1)}\;,
\end{equation}
which are complementary to the already existing elements in (\ref{kin-meas-2}). 

The analogous transformation has to be done for the connection forms. Adopting the notation of (\ref{coord-frame}), we 
separate the forms into two groups
\begin{equation}\label{con-sep}
\Sigma_m\,:\quad\underbrace{\omega_{1,n}^{(m)}\wedge \ldots \wedge \omega_{n-m,n}^{(m)}}_{\text{tangential}}\wedge 
\underbrace{\omega_{n-m+1,n}^{(m)}\wedge\ldots \wedge \omega_{n-1,n}^{(m)}}_{\text{normal}}\;.
\end{equation}
Again, the tangential directions transform by the coset elements of $S^{m-1}$
\begin{equation}
\begin{split}
\omega_{\alpha, n}^{(m)} &= x_m^{(m)}\,\omega_{\alpha, n-m+1}^{(1)} + \mathcal{O}(\omega_{\alpha,\beta}^{(1)})\\
&\hspace{3em}\text{for}\quad 
1\leq \alpha \leq n-m\;,\;\; n-m+2\leq\beta\leq n-1\;,
\end{split}
\end{equation}
whose products provide the complementary part to the $\text{SO}(n-k)$ form of (\ref{kin-meas-2})
\begin{equation}
\bigwedge_{m=2}^k\;\bigwedge_{\alpha=1}^{n-m}\omega_{\alpha,n}^{(m)} = \bigl[\prod_{m=2}^k (x_m^{(m)})^{n-m}\bigr]
\bigwedge_{m=2}^k\;\bigwedge_{\alpha=1}^{n-m}\omega_{\alpha, n-m+1}^{(1)}\;.
\end{equation}

On the other hand, the normal components of the connection forms (\ref{con-sep}) transform under the adjoint of  
$g_{\alpha\beta}\in \text{SO}(m)$
\begin{equation}
\begin{split}
\omega_{\alpha,n}^{(m)}& = g^{-1}_{\alpha\beta}\hat e_\beta^{(1)}\,d(g_{n \gamma}\hat e_\gamma^{(1)})
= g^{-1}_{\alpha\beta}d g_{n\beta} + \mathcal{O}(\omega_{\beta\gamma}^{(1)})\\
&\hspace{8em} \text{for}\quad n-m+1\leq \beta < \gamma \leq n-1
\end{split}
\end{equation}
with the coset element $g_{n \alpha}\in \text{SO}(m)/\text{SO}(m-1)$ leaving the normal vectors invariant. Forming 
their product, they combine into the group measure of $\text{SO}(k)$ modulo further connections
\begin{equation}
\begin{split}
\bigwedge_{m=2}^k \bigwedge_{\alpha=n-m+1}^{n-1} \omega_{\alpha,n}^{(m)} 
&= \bigwedge_{m=2}^k [d\text{vol}(S^{m-1}) + \mathcal{O}(\omega_{\beta\gamma}^{(1)})]\\
&\quad\;\;\;\, = d\text{vol}(\text{SO}(k)) + \mathcal{O}(\omega_{\beta\gamma}^{(1)})\;.
\end{split}
\end{equation}

Finally, inserting the transformed normal components into the kinematic measure (\ref{kin-meas-2}), yields a 
differential form which is obviously $\text{SO}(k)$ invariant
\begin{equation}
\begin{split}
&\int K(\Sigma_1\cap\ldots\cap\Sigma_k)\wedge K(D_2)\wedge\ldots\wedge K(D_k)\\
&\quad = \int \frac{\det{(GBh)}}{|\text{SO}(n-k)|} \prod_{m=2}^k \bigl(x_m^{(m)}\bigr)^{n-m+1} \, 
d\text{vol}(\text{SO}(k))
\wedge K(\Sigma_1)\wedge\ldots\wedge K(\Sigma_k)\;.
\end{split}
\end{equation}
The product of $x_m^{(m)}$ terms is the Jacobian $J$ of the transformation into spherical coordinates and has the 
parameterization in terms of $\text{SO}(m)$ rotation angles for $m=1,\ldots, k$
\begin{equation}
x_m^{(m)} = \prod_{\alpha=1}^{m-1} \sin{(\phi_{\alpha,m})}\;,
\end{equation}
proving the invariance of the weight functions (\ref{weights-2}) under $\text{SO}(k)$-generated permutations of 
particles indices and thus of Lemma~\ref{cor:weights}.
\bibliographystyle{spmpsci}
\bibliography{fmt_in_n_dimensions}

\providecommand{\noopsort}[1]{}\providecommand{\singleletter}[1]{#1}%
\begin{thebibliography}{10}
\providecommand{\url}[1]{{#1}}
\providecommand{\urlprefix}{URL }
\expandafter\ifx\csname urlstyle\endcsname\relax
  \providecommand{\doi}[1]{DOI~\discretionary{}{}{}#1}\else
  \providecommand{\doi}{DOI~\discretionary{}{}{}\begingroup
  \urlstyle{rm}\Url}\fi

\bibitem{schmidt-mixtures}
Brader, J.M., Esztermann, A., Schmidt, M.: Colloidal rod-sphere mixtures:
  Fluid-fluid interfaces and the {Onsager} limit.
\newblock Phys. Rev. E \textbf{66}(3), 031,401 (2002)

\bibitem{chern-1}
Chern, S.S.: On the kinematic formula in the euclidean space of n dimensions.
\newblock Am. J. Math. \textbf{74}, 227--236 (1952)

\bibitem{chern-2}
Chern, S.S.: Integral formulas for hypersurfaces in euclidean space and their
  applications to uniqueness theorems.
\newblock Indiana Univ. Math. J. \textbf{8}, 947--955 (1959)

\bibitem{chern-3}
Chern, S.S.: On the kinematic formula in integral geometry.
\newblock J. Math. Mech. \textbf{16}, 101--118 (1966)

\bibitem{gross}
Dreizler, R., Gross, E.: Density Functional Theory.
\newblock Springer (1990)

\bibitem{evans-dft}
Evans, R.: The nature of the liquid-vapour interface and other topics in the
  statistical mechanics of non-uniform, classical fluids.
\newblock Adv. Phys. \textbf{28}, 143--200 (1979)

\bibitem{mecke-fmt}
Hansen-Goos, H., Mecke, K.: Fundamental measure theory for inhomogeneous fluids
  of nonspherical hard particles.
\newblock Phys. Rev. Lett. \textbf{102}(1), 018,302 (2009)

\bibitem{mecke-fmt2}
Hansen-Goos, H., Mecke, K.: Tensorial density functional theory for
  non-spherical hard-body fluids.
\newblock J. Phys.: Condens. Matter. \textbf{22}, 364,107 (2010)

\bibitem{white-bear-2}
Hansen-Goos, H., Roth, R.: Density functional theory for hard-sphere mixtures:
  The white-bear version mark {II}.
\newblock J. Phys.: Condens. Matter \textbf{18}, 8413--8425 (2006)

\bibitem{helgason}
Helgason, S.: Differential Geometry, Lie Groups, and Symmetric Spaces.
\newblock Academic Press (1978)

\bibitem{hsiung}
Hsiung, C.C.: A First Course in Differential Geometry.
\newblock John Wiley (1981)

\bibitem{isihara-orig}
Isihara, A.: Determination of molecular shape by osmotic measurement.
\newblock J. Chem. Phys. \textbf{18}(11), 1446--1449 (1950)

\bibitem{leroux}
Kaouche, A., Leroux, P.: Mayer and {Ree-Hoover} weights of infinite families of
  2-connected graphs.
\newblock Seminaire Lotharingien de Combinatoire \textbf{61 A}, B61Af (2009)

\bibitem{kihara-2}
Kihara, T.: The second virial coefficient of non-spherical molecules.
\newblock J. Phys. Soc. Japan \textbf{6}(5), 289--296 (1951)

\bibitem{kihara-1}
Kihara, T.: Virial coefficients and models of molecules in gases.
\newblock Rev. Mod. Phys. \textbf{25}, 831--843 (1953)

\bibitem{korden-3}
Korden, S.: Beyond the {Rosenfeld} functional: Loop contributions in
  fundamental measure theory (2012)

\bibitem{korden-2}
Korden, S.: Deriving the {Rosenfeld} functional from the virial expansion.
\newblock Phys. Rev. E \textbf{85}(4), 041,150 (2012)

\bibitem{marechal-1}
Marechal, M., Goetze, H.H., H{\"a}rtel, A., L{\"o}wen, H.: Inhomogeneous fluids
  of colloidal hard dumbbells: Fundamental measure theory and monte carlo.
\newblock J. Chem. Phys \textbf{135}, 234,510 (2011)

\bibitem{m_k_m}
Marechal, M., Korden, S., Mecke, K.: Deriving fundamental measure theory:
  Reconciling the zero-dimensional limit with the virial approach, in
  preparation

\bibitem{marechal-3}
Marechal, M., L{\"o}wen, H.: Density functional theory for hard polyhedra.
\newblock Phys. Rev. Lett. \textbf{110}, 137,801 (2013)

\bibitem{marechal-2}
Marechal, M., Zimmermann, U., L{\"o}wen, H.: Freezing of parallel hard cubes
  with rounded edges.
\newblock J. Chem. Phys. \textbf{136}(14), 144,506 (2012)

\bibitem{mcdonald}
McDonald, I.R., Hansen, J.P.: Theory of Simple Liquids.
\newblock University of Cambridge (2008)

\bibitem{minkowski}
Minkowski, H.: Volumen und oberfl{\"a}che.
\newblock Math. Ann. \textbf{57}, 447--495 (1903)

\bibitem{rh-1}
Ree, F.H., Hoover, W.G.: Fifth and sixth virial coefficients for hard spheres
  and hard disks.
\newblock J. Chem. Phys. \textbf{40}, 939 (1964)

\bibitem{rh-3}
Ree, F.H., Hoover, W.G.: Reformulation of the virial series of classical
  fluids.
\newblock J. Chem. Phys \textbf{41}, 1635 (1964)

\bibitem{rh-2}
Ree, F.H., Hoover, W.G.: Seventh virial coefficients for hard spheres and hard
  disks.
\newblock J. Chem. Phys \textbf{46}, 4181 (1967)

\bibitem{rh-4}
Ree, F.H., Keeler, R.N., McCarthy, S.L.: Radial distribution function of hard
  spheres.
\newblock J. Chem. Phys \textbf{44}, 3407 (1966)

\bibitem{scaled-particle-1}
Reiss, H., Frisch, H.K., Lebowitz, J.L.: Statistical mechanics of rigid
  spheres.
\newblock J. Chem. Phys. \textbf{31}(2), 369--380 (1959)

\bibitem{riddell}
Riddell, R.J., Uhlenbeck, G.E.: On the theory of the virial development of the
  equation of the state of monoatomic gases.
\newblock J. Chem. Phys. \textbf{21}, 2056 (1953)

\bibitem{rosenfeld-structure}
Rosenfeld, Y.: Scaled field particle theory of the structure and the
  thermodynamics of isotropic hard particle fluids.
\newblock J. Chem. Phys. \textbf{89}(7), 4272--4287 (1988)

\bibitem{rosenfeld-freezing}
Rosenfeld, Y.: Free-energy model for the inhomogeneous hard-sphere fluid
  mixture and density-functional theory of freezing.
\newblock Phys. Rev. Lett. \textbf{63}(9), 980--983 (1989)

\bibitem{rosenfeld2}
Rosenfeld, Y.: Density functional theory of molecular fluids: Free-energy model
  for the inhomogeneous hard-body fluid.
\newblock Phys. Rev. E \textbf{50}(5), R3318--R3321 (1994)

\bibitem{rosenfeld-gauss2}
Rosenfeld, Y.: Free energy model for the inhomogeneou hard-body fluid:
  Application of the {Gauss-Bonnet} theorem.
\newblock Mol. Phys. \textbf{86}, 637--647 (1995)

\bibitem{rosenfeld-mixture}
Rosenfeld, Y., Levesque, D., Weis, J.: Free-energy model for the inhomogeneous
  hard-sphere fluid mixture - triplet and higher-order direct
  correlation-functions in dense fluids.
\newblock J. Chem. Phys. \textbf{92}(11), 6818--6832 (1990)

\bibitem{cros-ros-1}
Rosenfeld, Y., Schmidt, M., L{\"o}wen, H., Tarazona, P.: Dimensional crossover
  and the freezing transition in density functional theory.
\newblock J. Phys.: Condens. Matter \textbf{8}, L577--L581 (1996)

\bibitem{cros-ros-3}
Rosenfeld, Y., Schmidt, M., L{\"o}wen, H., Tarazona, P.: Fundamental-measure
  free-energy density functional for hard-spheres: Dimensional crossover and
  freezing.
\newblock Phys. Rev. E \textbf{55}, 4245--4263 (1997)

\bibitem{roth-rev}
Roth, R.: Fundamental measure theory for hard-sphere mixtures: A review.
\newblock J. Phys.: Condens. Matter \textbf{22}, 063,102--063,120 (2010)

\bibitem{white-bear-1}
Roth, R., Evans, R., Lang, A., Kahl, G.: Fundamental measure theory for
  hard-sphere mixtures revisited: The white bear version.
\newblock J. Phys.: Condens. Matter \textbf{14}, 12,063--12,078 (2002)

\bibitem{santalo-book}
Santalo, L.A.: Integral Geometry and Geometric Probability.
\newblock {Addison-Wesley} (1976)

\bibitem{santos-2}
Santos, A.: Class of consistent fundamental-measure free energies for
  hard-sphere mixtures.
\newblock Phys. Rev. E \textbf{86}, 040,102 (2012)

\bibitem{santos-1}
Santos, A.: Note: An exact scaling relation for truncatable free energies of
  polydisperse hard-sphere mixtures.
\newblock J. Chem. Phys. \textbf{136}, 136,102 (2012)

\bibitem{schmidt-dft}
Schmidt, M.: Fluid structure from density-functional theory.
\newblock Phys. Rev. E \textbf{62}(4), 4976--4981 (2000)

\bibitem{weil}
Schneider, R., Weil, W.: Stochastic and Integral Geometry.
\newblock Springer (2008)

\bibitem{tarazona}
Tarazona, P.: Density functional for hard sphere crystals: A fundamental
  measure approach.
\newblock Phys. Rev. Lett. \textbf{84}, 694--697 (2000)

\bibitem{tarazona-rosenfeld}
Tarazona, P., Rosenfeld, Y.: From zero-dimensional cavities to free-energy
  functionals for hard disks and hard spheres.
\newblock Phys. Rev. E \textbf{55}, R4873--R4876 (1997)

\bibitem{uhlenbeck-ford-1}
Uhlenbeck, G.E., Ford, G.W.: The Theory of Linear Graphs with Applications to
  the Theory of the Virial Development of the Properties of Gases,
  \emph{Studies in Statistical Mechanics}, vol.~1.
\newblock Interscience Publishers (1962)

\bibitem{wertheim-1}
Wertheim, M.S.: Fluids of hard convex molecules {I.} basic theory.
\newblock Mol. Phys. \textbf{83}, 519--537 (1994)

\bibitem{wertheim-2}
Wertheim, M.S.: Fluids of hard convex molecules {II.} two-point measures.
\newblock Mol. Phys. \textbf{89}, 989--1004 (1996)

\bibitem{wertheim-3}
Wertheim, M.S.: Fluids of hard convex molecules {III.} the third virial
  coefficient.
\newblock Mol. Phys. \textbf{89}, 1005--1017 (1996)

\bibitem{wertheim-4}
Wertheim, M.S.: Third virial coefficient of hard spheroids.
\newblock Mol. Phys. \textbf{99}, 187--196 (2001)

\end{thebibliography}
\end{document}